\documentclass[aps,final,authoryear]{revtex4-2}

\usepackage{balance}
\usepackage{times,mathptmx}
\usepackage{graphicx} 

\usepackage[english]{babel}
\addto{\captionsenglish}{
  \renewcommand{\refname}{Notes and references}
}
\usepackage{array}
\usepackage{xcolor}
\usepackage[T1]{fontenc}
\usepackage{setspace}
\usepackage[compact]{titlesec}
\usepackage{hyperref}
\usepackage{subfigure}

\usepackage{epstopdf}

\definecolor{cream}{RGB}{222,217,201}

\begin{document}


\title{Slippery flows of a Carbopol gel in a micro-channel}


\author{Eliane Younes}
\email[Eliane Younes]{Eliane.Younes@univ-nantes.fr}
\affiliation{Universit\'{e} de Nantes, CNRS, Laboratoire de thermique et \'{e}nergie de Nantes, LTeN, F-44000 Nantes, France}

\author{Volfango Bertola}
\email[Volfango Bertola]{Volfango.Bertola@liverpool.ac.uk}
\affiliation{Laboratory of Technical Physics, School of Engineering, University of Liverpool, Liverpool, L69 3GH, United Kingdom}

\author{Cathy Castelain}
\email[Cathy Castelain]{Cathy.Castelain@univ-nantes.fr}
\affiliation{Universit\'{e} de Nantes, CNRS, Laboratoire de thermique et \'{e}nergie de Nantes, LTeN, F-44000 Nantes, France}

\author{Teodor Burghelea}
\email[Teodor Burghelea]{Teodor.Burghelea@univ-nantes.fr}
\affiliation{Universit\'{e} de Nantes, CNRS, Laboratoire de thermique et \'{e}nergie de Nantes, LTeN, F-44000 Nantes, France}


\date{\today}

\begin{abstract}
The ability to predict and/or control wall slip is a fundamental problem in the hydrodynamics of yield stress fluids, which is poorly understood to date and has important applications in bio- and micro-fluidic systems. Systematic measurements of steady flows of a simple yield stress fluid (Carbopol Ultrez 10) in a plane acrylic micro-channel are used to establish rigorous scaling laws for the wall velocity gradient and the slip velocity.
By means of epi-fluorescent microscopy combined with a custom developed Digital Particle Image Velocimetry (DPIV) technique, times series of velocity fields were measured within a wide range of flow rates and three distinct flow regimes were identified: full plug, partial plug and fully yielded. Corresponding to each flow regime, wall velocity gradients and slip velocities were obtained by extrapolating the velocity profiles using a smoothing spline function. By combining the flow field measurements with the macro-rheological measurements, scaling laws for the wall velocity gradient and the slip velocity with the wall shear stress were identified, and compared with results from the literature. Detailed microscopic measurements of the velocity field enabled an assessment of the effectiveness of a chemical treatment of the channel walls meant to suppress wall slip  proposed by Metivier and coworkers in Ref. \cite{StickSlip}.
\end{abstract}


\maketitle
\tableofcontents
\listoffigures

\clearpage
An ubiquitous phenomenon observed during flows of yield stress materials along smooth solid surfaces is the slip of the velocity near the wall. From a phenomenological standpoint it is believed that the wall slip originates from the presence of a depleted layer of solvent near the solid surface and the bulk of the material \emph{"slides"} over it.  However, to our best knowledge, this phenomenological picture has not been verified yet by means of direct microscopic visualisation of the material microstructure in the vicinity of the solid wall. As compared to the case of slip during flows of molecular (simple, or Newtonian) fluids where, based on the precise knowledge of the intermolecular forces, a rigorous theoretical framework may be derived \cite{TABELING2004531,PhysRevLett.92.166102}, the wall slip of yield stress materials remains elusive. A key step in understanding the physics of the wall slip in yield stress fluids is to reliably measure scaling laws of the slip velocity with respect to the wall shear stresses and the wall velocity gradients. The scaling behavior during flows of yield stress fluids in the presence of wall slip remains, however, an open topic. To substantiate this statement, we present in the following a brief literature review.
The existing studies of wall slip during flows of yield stress fluids along smooth solid surfaces may be grouped in three distinct classes. 

A first class refers to either phenomenological modelling and or numerical studies of the wall slip phenomenon, \cite{R7,DAMIANOU201613}. Such studies typically rely on assumptions concerning the micro-structure of the material in vicinity of the solid wall which are often difficult to probe by means of direct visualisation. As the numerical studies are concerned, the evolution of the velocity gradients and stresses near the wall is typically obtained by assuming a known bulk rheological behaviour of material.  
Depending on the phenomenological assumptions being made, various slip laws are derived. The agreement between various phenomenological scaling predictions is, to our best understanding, not very satisfactory in the sense that various authors obtain different scaling results even when the same material is concerned which rarely match the experimental results. 

A second and significantly broader class of studies refers to macroscopic experiments performed in either rheometric \cite{R2,R3,R6,R8,R9,R18,R19} or non-rheometric \cite{unsteady,RheoPIV} geometries. Such studies are sometimes combined with rather simple phenomenological models and scaling laws are obtained. 
The presence of wall slip complicates the determination of the yield stress \cite{kaylonslip} and systematically biases the assessment of the solid-fluid transition \cite{volfangoslip,unsteady}. 

A systematic experimental study of the wall slip phenomenon in a rheometric flow was reported by Meeker and coworkers, \cite{R2}. By assessing the slip behaviour via macro-rheological tests combined with measurements of the velocity geometry within the gap of the rheometric geometry, they identify three distinct slip regimes. They observed wall slip for applied stresses close to the yield stress $\tau_y$ whereas in a fully yielded regime this effect was negligible.
A second macro-rheological study of the wall slip observed during flows of hard-sphere colloidal glasses was reported by Ballesta and coworkers, \cite{R6}. This study highlights only two slip regimes: solid (\textbf{S}) and fluid (\textbf{F}). Within the regime \textbf{(S)} a Bingham type slip regime is observed whereas within the \textbf{(F)} regime a Herschel-Bulkley slip behaviour is observed. 

The experimental study by Meeker and coworkers \cite{R2} is complemented by a phenomenological model based on micro-elasto-hydrodynamic lubrication picture which allows one to derive scaling laws for the full slip case. The main assumption behind this phenomenological approach is the existence of a depleted layer of solvent of width $\delta$ in the vicinity of the solid surface. However, the authors note that \emph{"the slip layer is not resolvable in our setup, indicating that its thickness is smaller than $50 \mu m$"} meaning that during their experiments this assumption could not be probed by direct observation. A similar estimate of the layer thickness has been reported by Zhang and coworkers, \cite{R9}.
Such  small values explain  why a direct visualisation of the depleted layer of solvent has never been reported in the literature.

%

A second sub-class of experimental macroscopic studies refers to non-rheometric flows such as laminar pipe flows.  A study of the flow in the presence of wall slip in a glass-made capillary tube of a $0.2~wt\%$ Carbopol gel was performed by P\'{e}rez-Gonz\'{a}lez et al. \cite{RheoPIV}. By combining the Particle Image Velocimetry technique with classical rheological measurements they found  that the solid-fluid transition occurs gradually which complicates the determination of the yield stress.  The authors estimate the wall velocity gradients using the relation between the imposed flow rate $Q$ and the analytical solution of a fully developed Poiseuille flow for a Newtonian fluid. 
An experimental study of a laminar flow of a Carbopol gel in a glass made cylindrical pipe with a radius $R=1.9~mm$ was performed by Poumaere and his coworkers \cite{unsteady}.  Several important findings of this study can be summarised as follows.  First, consistently with the findings of P\'{e}rez-Gonz\'{a}lez and coworkers, the transition from a solid like to a fluid like regime is smooth and mediated by an intermediate regime where the solid and fluid behaviors coexist. Second, the wall slip effects were found to be more pronounced around the yield point. 
 
 A third class of studies refers to microscopic flows where the influence of the confinement is expected to play an important role and can not be de-coupled from the wall slip and yielding behaviour, \cite{LIU201825,Geraud2013,R17,barentin}. 
As compared to the second class, this third class of studies is somewhat less represented although the microscopic wall slip behaviour is relevant to a number of microscopic experimental settings such as the impact and/or atomisation of viscoplastic drops \cite{Bertola2009,German201018,German20101,cism} with a number of applications in fuel industry and microscopic blood flows \cite{doi:10.1063/1.5011373,thesisv}. 

Geraud et al. \cite{Geraud2013} studied the influence of confinement on the flow of a Carbopol microgel. They compare the bulk rheological behavior assessed via a macroscopic rheological tests and the velocity profiles measured in rough micro-channels. They found a strong disagreement between the bulk prediction of the velocity profiles and the ones measured in the micro-channels and they attribute this discrepancy to the confinement of the flow.
 Liu et al.  \cite{LIU201825} compared the measured velocity profiles in micro-channels to simulations performed based on the bulk rheology. They show that the confinement effects become important when the dimensions of the channel are comparable to the characteristic size of the material's microstructure and, consequently, the measured velocity profiles differ from those computed numerically using the Herschel-Bulkley constitutive relation.  These two studies address the effects of the confinement on the flow behavior but do not directly tackle the wall slip phenomenon.  

Most recently, P\'{e}m\'{e}ja and coworkers have presented a systematic experimental investigation of wall slip regimes of a Carbopol gel in a micro-channel, \cite{barentin}. As in Ref.  \cite{RheoPIV} the authors compute the wall shear stress using the pressure drop. To rationalise their results, the authors use a phenomenological model similar to the one proposed by Meeker and  coworkers in Ref. \cite{R2}. 
Based on measurements of the transversal profiles of the axial velocity, they identify two distinct scaling regimes as the driving pressure drops are gradually increased. 
Seth and coworkers studied the slip behaviour in a microscopic rotational shear cell (Linkam
CSS 450)  by means of Particle Tracking Velocimetry, \cite{R11}. They identify two slip regimes separated by an abrupt transition. For stresses exceeding the yield stress $\tau>\tau_y$ the slip velocity scales linearly with the stress and the slope of this linear dependence does not depend on the chemistry of the solid surface. In the plug flow regime ($\tau<\tau_y$ ), the slip velocity measured at an imposed stress is very sensitive to the
nature of the surfaces. To rationalise their findings, the authors use an elasto-hydrodynamic lubrication model which allows them to compute the wall shear stress as a function of the slip velocity and the thickness $\delta$ of the lubricating layer.
A microscopic study of the slip behaviour of a soft glassy material is presented by Mansard and coworkers, \cite{R12}. They highlighted a strong dependence of the slip behaviour on the degree of roughness of the solid surface which was accurately controlled by means of micro-patterning using standard micro-photolithography procedure.

An important conclusion that can be drawn from the existing experimental studies performed in macroscopic or microscopic geometries is that the wall slip behaviour can not be de-coupled from the yielding behaviour. Unlike in the case of the polymer melts where the wall slip emerges at large rates of strain \cite{savasslip,HATZIKIRIAKOS2012624}, in the yield stress materials the picture is strikingly different. All the  experiments reviewed above indicate that the wall slip effects are more pronounced in a range of low shear rates (where the solid-fluid transition takes place) and are significantly less important far beyond the yield point when a simple constitutive equation (such as the Herschel-Bulkley constitutive law) is reliable, \cite{volfangoslip,unsteady}. Whereas about this fact there seems to be a general consensus within the viscoplastic community, various authors use different assumptions (which in most of the cases could not be verified by means of direct visualisation of the flow in windows sufficiently small and close to the solid boundary) and phenomenological models (e.g. the elasto-hydrodynamic lubrication model sometimes combined with yet other assumptions on the nature of intermolecular forces responsible for the wall slip) in order to rationalise their findings and obtain scaling laws. 
An additional drawback of some of the previous studies relates to using questionable correlations (e.g. computing the wall shear rate using the the flow rate or pre-assuming a Herschel-Bukley like velocity profile which, as shown in Ref. \cite{LIU201825}) is not accurate in the presence of confinement.
The overall conclusion is that, in spite of the reliability of the macroscopic scale experiments in capturing the wall slip behaviour, there seems to exist no universal consensus on neither the scaling behaviour nor the fundamental physical mechanisms responsible for the wall slip behaviour.

The central aim of the present contribution is to provide a systematic experimental characterization of the microscopic wall slip phenomenon in terms of scaling laws of the slip velocity with both the wall shear stresses and the wall velocity gradients. The novelty of the approach we propose in the following consists of assessing these scaling laws by judiciously correlating microscopic measurements of the wall velocity gradients performed in a plane micro-channel with the bulk rheological behaviour assessed by macro-rheological measurements performed in a steady state. By doing so we avoid making any of the assumptions other authors made such as the presence of a lubricating layer and/or a constitutive relation and/or the relationship between the wall shear stresses and the driving pressures. The study is complemented by a investigation of the effectiveness of a chemical treatment of the channel walls meant to suppress the wall slip.  

The paper is organised as follows. The rheological characterization of the working fluid is presented in Sec. \ref{sec:Rheology}. The design of micro-channel setup as well as the data acquisition protocol are presented in Sec. \ref{subsec:MicrochannelDesign}. The data analysis procedure is discussed in Sec. \ref{subsec:PIV measurements}. A systematic characterization of the velocity profiles  is presented in Sec.  \ref{sec:results}. By combining the micro-fluidic and macro-rheological measurements, we identify in Sec. \ref{sec:results_wallslipbehavior} different scaling laws of the wall slip phenomenon corresponding to different flow regimes. A detailed comparison of the experimentally found scaling laws with the results from the literature is presented in Sec. \ref{sec:comparison}. 
In Sec. \ref{sec:inhibitting_wallslip} we explore the possibility of inhibiting the wall slip via a chemical treatment of the micro-channels. 
The paper closes with a summary of the main conclusions and a discussion of several perspectives this study may open, Sec. \ref{sec:conclusion}.

\section{ Experimental methods}\label{sec:experimental}

\subsection{Choice, preparation and rheological characterization of the working fluid}\label{sec:Rheology}

We chose a Carbopol gel as a working fluid. The rational behind this choice is the following. First, such gels are optically transparent which allows an in-situ visualisation of their flows and subsequent measurements of the flow fields. Second, they are chemically stable over extended  periods of time.
Third, and foremost, the Carbopol gels have been considered as \emph{``model''} yield stress materials for  over two decades \cite{model1,model2,piaucarbopol}: they practically exhibit no thixotropy and their deformation may be described by the Herschel-Bulkley law.
 
However, it has been shown recently that even in the case of a Carbopol gel, the solid-fluid transition does not occur at a well defined value of the applied stress but gradually (within a finite range of stresses and, when forced in an unsteady fashion, they equally exhibit weak thixotropic effects manifested through a rheological hysteresis which is related to the degree of steadiness of the forcing, \cite{manneville,solidfluid,softmatter2,thermo,miguelstab,cism}). In addition to that, the Herschel-Bulkley constitutive relationship can not reliably describe the deformation states within the transitional regime.

To avoid such time-dependent effects which may bias the slip behaviour, we focus through this paper on steady state microscopic flows which we correlate to steady state rheological measurements.

\subsubsection{Fluid preparation}\label{subsec:Fluid preparation}

The working fluid was an aqueous solution of Carbopol Ultrez 10 with the concentration of $0.1~wt\%$. 
Carbopol is the generic trade name of a cross-linked polyacrylic acid ($COO^{-}H^{+}$) with high molecular weight. In an anhydrous state, it is commercialized in the form of a white powder soluble in aqueous solvents. After the addition of a neurtralizing agent such as sodium hydroxide (NaOH), a clear gel is obtained. The Carbopol gels exhibit an elasto- viscoplastic rheological behavior in a neutral state due to the presence of a jammed spongy microstructure, \cite{debruyn2,debruynscaling,solidfluid}.

The Carbopol gel was prepared according to the following protocol. 
First, the right amount of anhydrous Carbopol was dissolved in deionized water using a magnetic stirring device at a speed of $1000~rpm$.
The degree of mixing/dissolution was assessed visually by monitoring the optical isotropy of the solution. 
Next, the $pH$ of the solution was gradually increased from $3.2$ to $7$ by gradual titration with a small amounts of a $10~wt\%$ aqueous $NaOH$ solution gradually pipetted while gently mixing the solution. The rheological tests were performed after seeding the Carbopol solution with the fluorescent tracers as detailed in Sec. \ref{subsec:MicrochannelDesign}.
The average density of the Carbopol solution was $\rho \approx 1100~kg.m^{-3}$.

\subsubsection{Rheological measurements}\label{subsec:Rheological measurments}
The rheological measurements were performed using a controlled stress rotational rheometer (Mars $III$, Thermofischer Scientific) equipped with a nano-torque module. Tests were performed using a parallel plate geometry with a diameter $D=35~mm$ and a gap $d=1~mm$. To prevent the wall slip, glass paper with an average roughness of $500~\mu m$ was glued on each plate. To account for the addition of the glass paper on the rotating plate of the device, the inertia of the device was recalibrated. The absence of any wall slip effect was verified by measuring flow curves in subsequent tests performed with several values of the gap and showing that all measurements perfectly overlap. To prevent the evaporation of the solvent during the rheological measurements a thin layer of commercial oil was added to the free meniscus of the sample. 

Rheological measurements were performed according to the following protocol. First, the sample was pre-sheared at a constant applied stress larger than the yield stress for $300~s$ and allowed to relax for another $300~s$. Then, to assess the rheological behavior of the Carbopol gel in different deformation regimes, a commonly used rheological test consisting of loading the material according to an increasing stress ramp was applied to a fluid sample. The duration of each step was  $t_0=5~s$ and the data averaging time per stress value was $\delta t_0=2~s$.

To test the reproducibility and quantitatively assess the instrumental error, each rheological measurement was repeated three times with a fresh sample. 

\begin{figure}[h]
\centering
{\includegraphics[height=6cm]{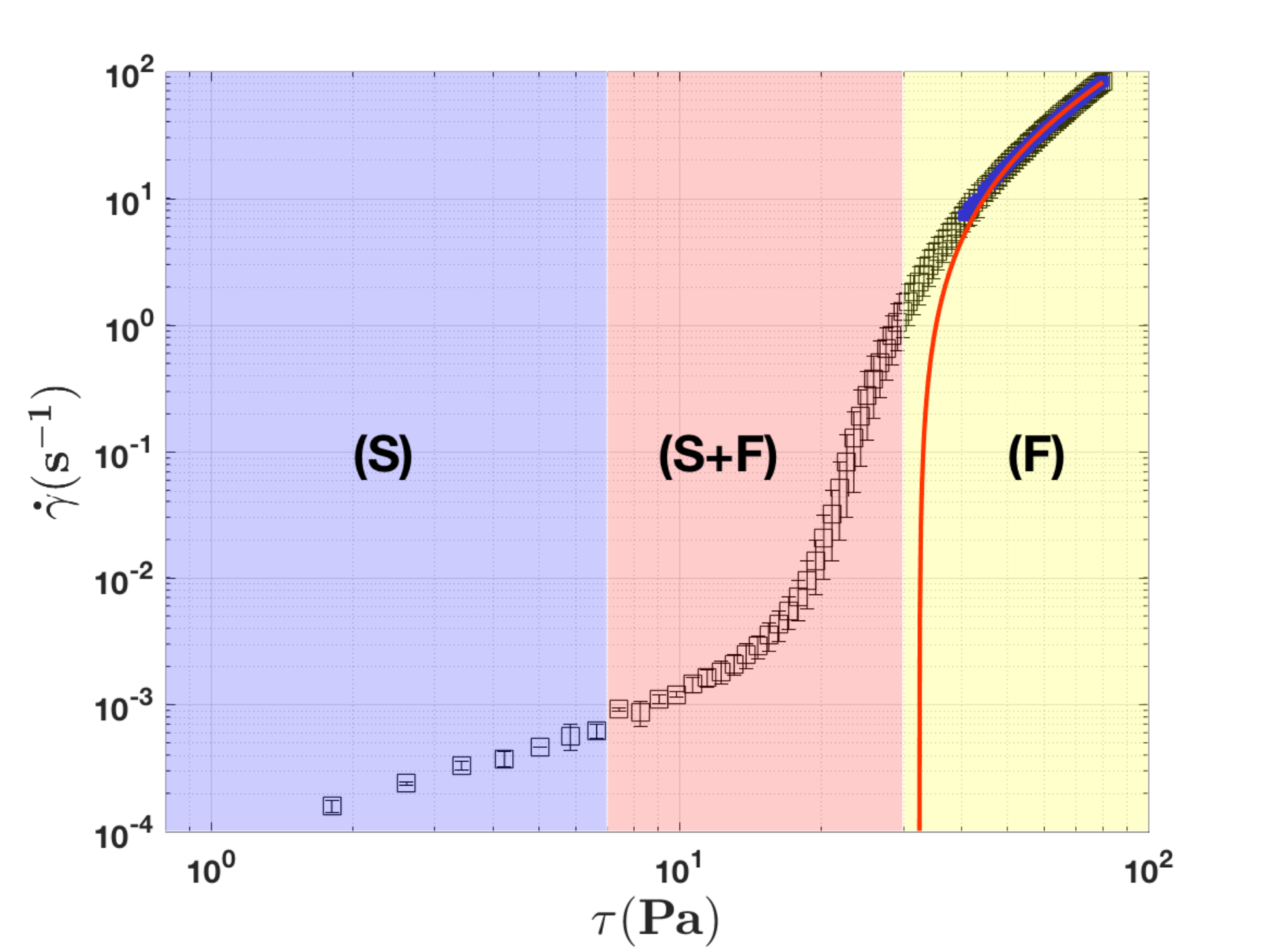}}
\caption{Dependence of the shear rate $\dot{\gamma}$ on the on the applied stress $\tau$. The red line represents the Herschel - Bulkley fit that gives $\tau_y=32.5 \pm 2.5~Pa$, $K=2.8 \pm 0.215~Pa.s^n$, $n=0.6403 \pm 0.0299$. The symbols marking the highlighted regions denote the deformation regimes and are explained in the text: \textbf{(S)} - solid, \textbf{(S + F)} - solid-fluid coexistence, \textbf{(F)} - fluid.
\label{F:flow_curve}}
\end{figure}

The rheological flow curve  is illustrated in Fig. \ref{F:flow_curve}. Three distinct deformation regimes consistent with previous macro rheological investigations, \cite{solidfluid,Sainudiin15,softmatter2,thermo,miguelstab,cism}  are observed. For low values of applied stress, the variation of the rate of material deformation $\dot\gamma$ with the applied stress is negligibly small. This corresponds to a solid-like behavior of the material, region \textbf{(S)} in Fig. \ref{F:flow_curve}. For high values of applied stress (beyond the yield stress), the material becomes fully yielded, region \textbf{(F)} in Fig. \ref{F:flow_curve}, and its behavior follows the Herschel-Bulkley fit, the red line in Fig. \ref{F:flow_curve}.

For intermediate values of the applied stresses, the solid- fluid transition is not direct and the behavior of the material corresponds neither to a solid-like regime nor to a viscous flow regime, region \textbf{(S+F)} in Fig. \ref{F:flow_curve}. Based on the Herschel-Bulkley fit, the yield stress of the working material is around $\tau_y \approx 32.5~Pa$.

A natural question one may ask at this point is to what extent the steady state rheological measurements presented in Fig. \ref{F:flow_curve} are relevant to the steady micro-channel flow under investigation. As shown in Ref. \cite{solidfluid,cism} for characteristic times per stress step $t_0$ larger than $2s$ (we recall that the rheological data was acquired with $t_0=5s$) the regimes \textbf{(F)} and \textbf{(S+F)} are practically independent of $t_0$ while a weak dependence is always observed in the regime $\textbf{(S)}$. This means that as long as we restrict ourselves to the regimes \textbf{(F)} and \textbf{(S+F)}, a meaningful correlation between the rheological measurements presented in Fig. \ref{F:flow_curve}  and the flow fields measured in the micro-channel can be made.     

\subsection{Micro-channel design, microscopic flow control and data acquisition protocol}\label{subsec:MicrochannelDesign}
\begin{figure}[h]
\centering
{\includegraphics[height=4cm]{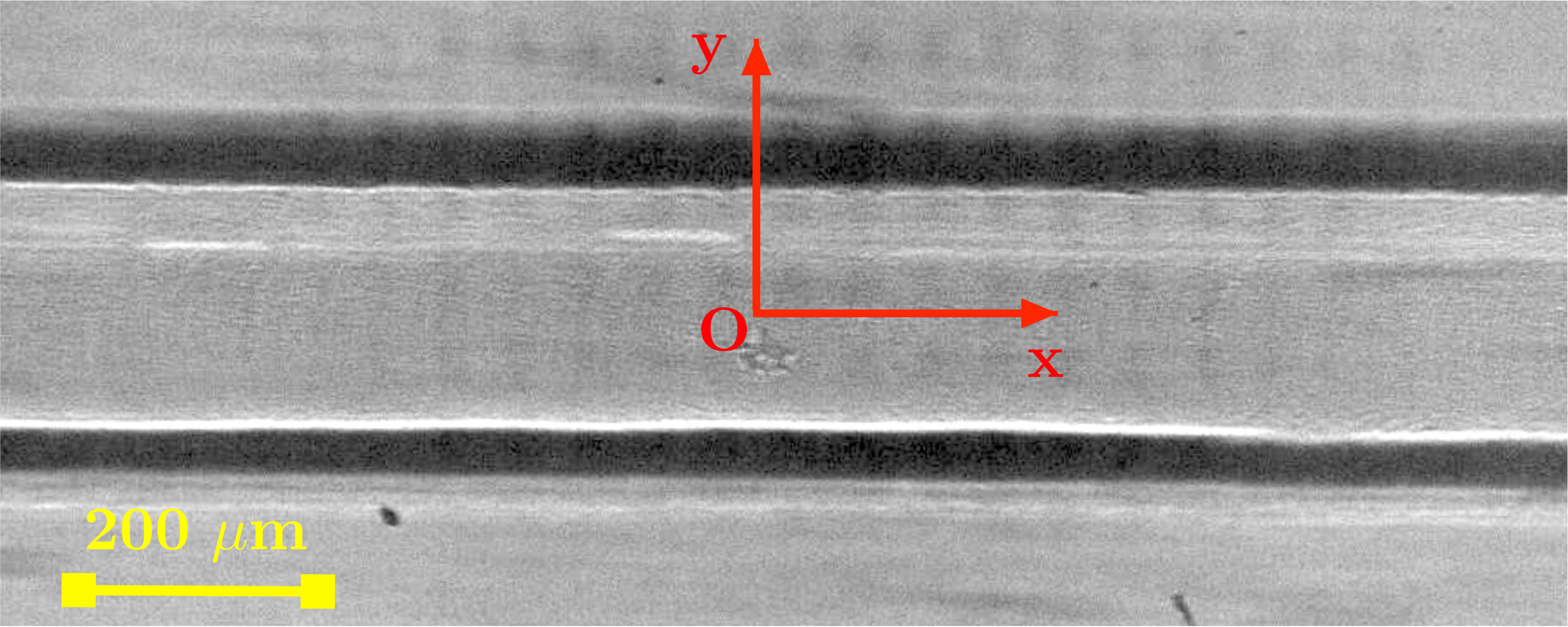}}
\caption{Micro-graph of the micro-channel. The image was acquired in white light.}
\label{F:Channel}
\end{figure}

The experiments have been performed in a straight micro-channel. 
The width of the micro-channel is $W=200~\mu m$, its depth is $H=200~\mu m$ and its length $L=4~cm$. A micrograph of the channel is presented in Fig. \ref{F:Channel}. 

\begin{figure}[h]
\centering
{\includegraphics[height=6cm]{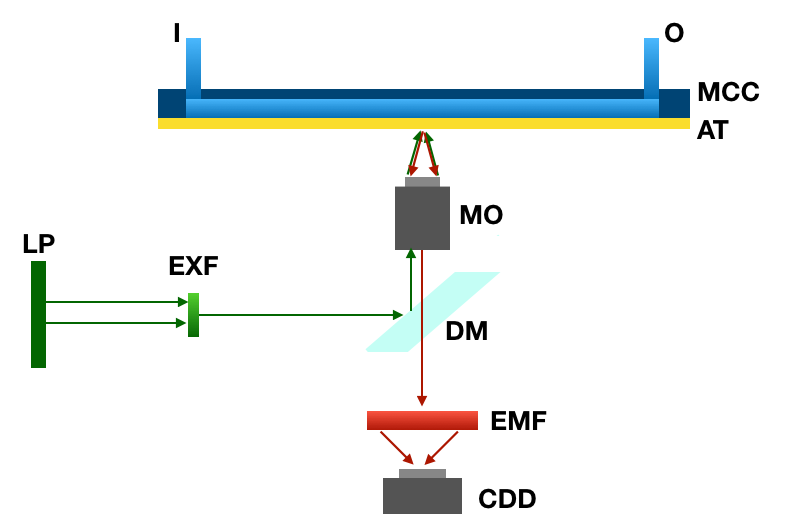}}
\caption{ Schematic view of the microfluidic experimental setup: MCC: micro-channel chip, I: micro-channel inlet, O: micro-channel outlet, AT: adhesive tape, MO: microscope objective, DM: dichroic mirror, LP: light emitting lamp, EXF: excitation filter, EMF: emission filter, CCD: digital camera.
\label{F:Setup}}
\end{figure}

The micro-channel was machined in an acrylic block with the dimensions $5~cm\times 3~cm \times 0.5~cm$ using a fast spinning ($14 000~rot/min$) micro-milling head (Nakanishi, model $HES 510 - BT40$) mounted on a commercial computer controlled milling machine (Twinhorn, model $VH1010$). 
By a precise alignment of the initial aluminium block on the stage of the milling machine, the depths of the micro-channel was uniform over its entire lengths with an end to end variation smaller than one percent. The average roughness of the edges of the micro-channel as resulted from the micro milling process is roughly of the order of a micron which accounts for half percent of the channel width. This point is fully supported by the micrograph of the micro-channel shown in Fig. \ref{F:Channel}.
We note that this level of edge smoothness is comparable to that obtained via the classical micro lithography techniques used to produce PDMS micro-channels, \cite{Thorsen580}. 
The micro-channel was sealed with a strong and optically transparent adhesive tape ($3M$, model $727-1280$). The micro-channel chip is mounted on an inverted epifluorescent microscope (Axio Observer A1, serial no3832002215. ), Fig. \ref{F:Setup}.

The microscopic flows have been visualised through a $20X$ magnification objective (Zeiss EC Plan-NEOFLUAR 
20x/0,5) with a numerical aperture $NA=0.35$ and a long working distance $WD=70~mm$. 
The microscopic flows were generated using a high precision micro-syringe pump (KdScientific, model Legato 110). To insure a steady flow rate $Q$ which was crucial for this study, we have used a $10~ml$ gas tight syringe (Hamilton, model 1010LT). The flows were illuminated with a powerful halogen lamp coupled to the inverted microscope.
To visualise the flow, the working fluid was seeded with a minute amount of buoyantly neutral fluorescent tracers with a diameter of $0.92~\mu m$ (Fluoresbrite Multifluorescent from Polysciences).
A series of $1000$ flow images was acquired corresponding to each value of the flow rate $Q$ with a digital camera. The images were acquired at the mid-plane of the micro-channel and at mid-distance downstream.
For the experiments where the flow rate did not exceed $10 ~\mu L/min$, we have used a Prosilica GE camera with $16$ bit quantisation (model GE680C from Allied Technologies). The maximal frame rate achievable with this camera is $200 ~fps$.
During the experiments with flow rates $Q \in [10, 23]~ \mu L/min$, a Mikrotron camera has been used up to a frame rate of $500~fps$.

\subsection{Data analysis}\label{subsec:PIV measurements}
\begin{figure}[h]
\centering
{\includegraphics[height=6cm]{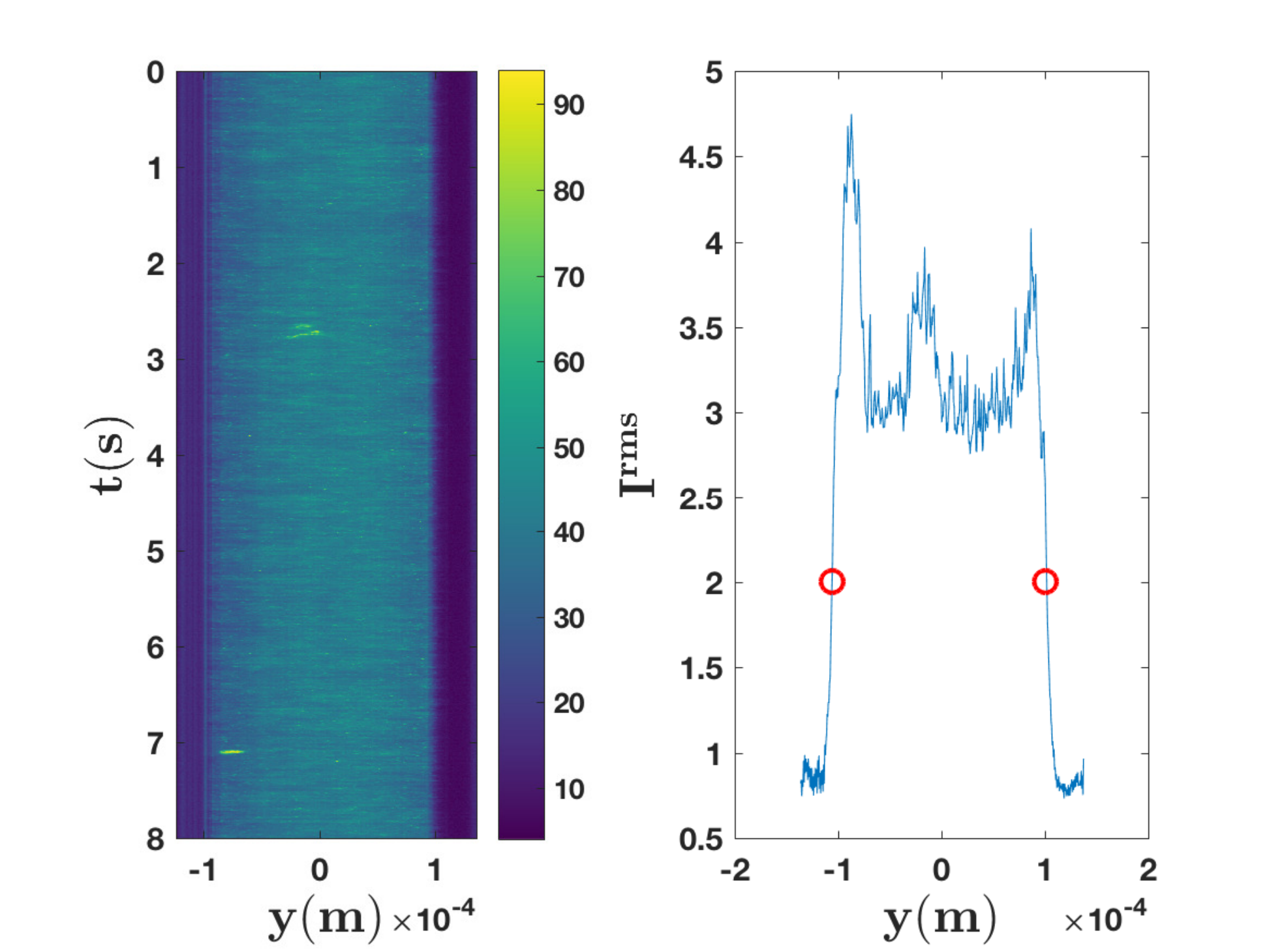}}
\caption{ The plot on the left refers to the space time of the transversal profiles of image brightness acquired at the middle of the field of view. The plot on the right refers to the root mean square deviation (rms) of the brightness of the space time diagram along the temporal direction.
\label{F:edges}}
\end{figure}

The main tool used to systematically characterize the microscopic flow was an adaptive multi-grid Digital Particle Image Velocimetry (DPIV) (see Refs. \cite{scarano,piv2} for a detailed description of the method) entirely developed in house under Matlab (together with the ``Image Processing Toolbox") which followed the steps briefly described below. 
The first step of the data acquisition procedure was to precisely identify the borders of the micro-channel. Though apparently trivial, this step is crucial for correctly measuring the slip velocity and the wall velocity gradients which represent the core of the present study.  
To do so, we have build from the entire sequence of acquired images a space time diagram $xt=xt(t,y)$ of the transversal (to the flow direction ) profiles of image brightness acquired at the middle of the field of view., Fig. \ref{F:edges}(a). 
Next, we have computed the root mean square deviation (rmsd) of the space-time diagram along its temporal axis according to:
\begin{equation}\label{eq:rms}
I^{rms}(y) = \left( \left< (xt(t,y)-\left< xt(t,y)\right>_t)^2\right>_t \right)^{1/2}
\end{equation}
Here $\left< \cdot \right>_t$ denotes the average of a two dimensional array computed along its temporal dimension $t$.
Last,  we identify the edges of the channel by the finding the two local maxima of the dependence $\left \vert \frac{d I^{rms}(y)}{dy}\right \vert$ represented in Fig. \ref{F:edges}(b) by the circles. 

Second, a background was calculated by averaging all the images of the first time series we have acquired. Third,  the average background was subtracted from images.  Finally, pairs of pre-processed images separated in time by the inter-frame $t_1$ are passed to  a multi-pass $DPIV$ algorithm using a sequence of squared interrogation windows with sizes $ \left[128,64,32,16,8 \right]$.   
As a post-processing step, each computed velocity field obtained from the DPIV algorithm was filtered using a median filter. The spatial resolution of the velocity fields was $4.5~\mu m$ ($44$ times smaller than the width of the channel) which suffices for a reliable calculation of the velocity gradients in the proximity of the channel walls. Corresponding to each driving flow rate, we have computed the average of $1000$ subsequent velocity fields together with their root mean square deviation (rms) which allowed one to specify error bars.  

To compute accurately the velocity gradients, each profile of the time averaged velocity was first fitted by a smoothing spline function which was differentiated analytically in order to avoid the inherent errors associated to the classical numerical differentiation. 

To extract the wall slip velocity, we have extrapolated each spline fitting function of the axial velocity and computed its value at the level of the wall. The velocity gradient near the wall was assessed via the slope of the last four points of the smoothing spline function. 
The size of the rigid plug located along the centre-line of the micro-channel was estimated by detecting the position where the transversal gradient of the axial velocity becomes larger than 0.3 $\%$ of its maximal value along the velocity profile. The plug velocity was calculated by averaging the velocity within the plug region. 

\section{Results} \label{sec:results}

\subsection{Description of the flow regimes in the presence of wall slip} \label{sec:results_flowregimes}
We focus in the following on the experimental characterization of the microscopic flow profiles of a Carbopol gel in a steady channel flow for several flow rates. 

\begin{figure}
\centering
\subfigure[]{
     \includegraphics [height=6cm] {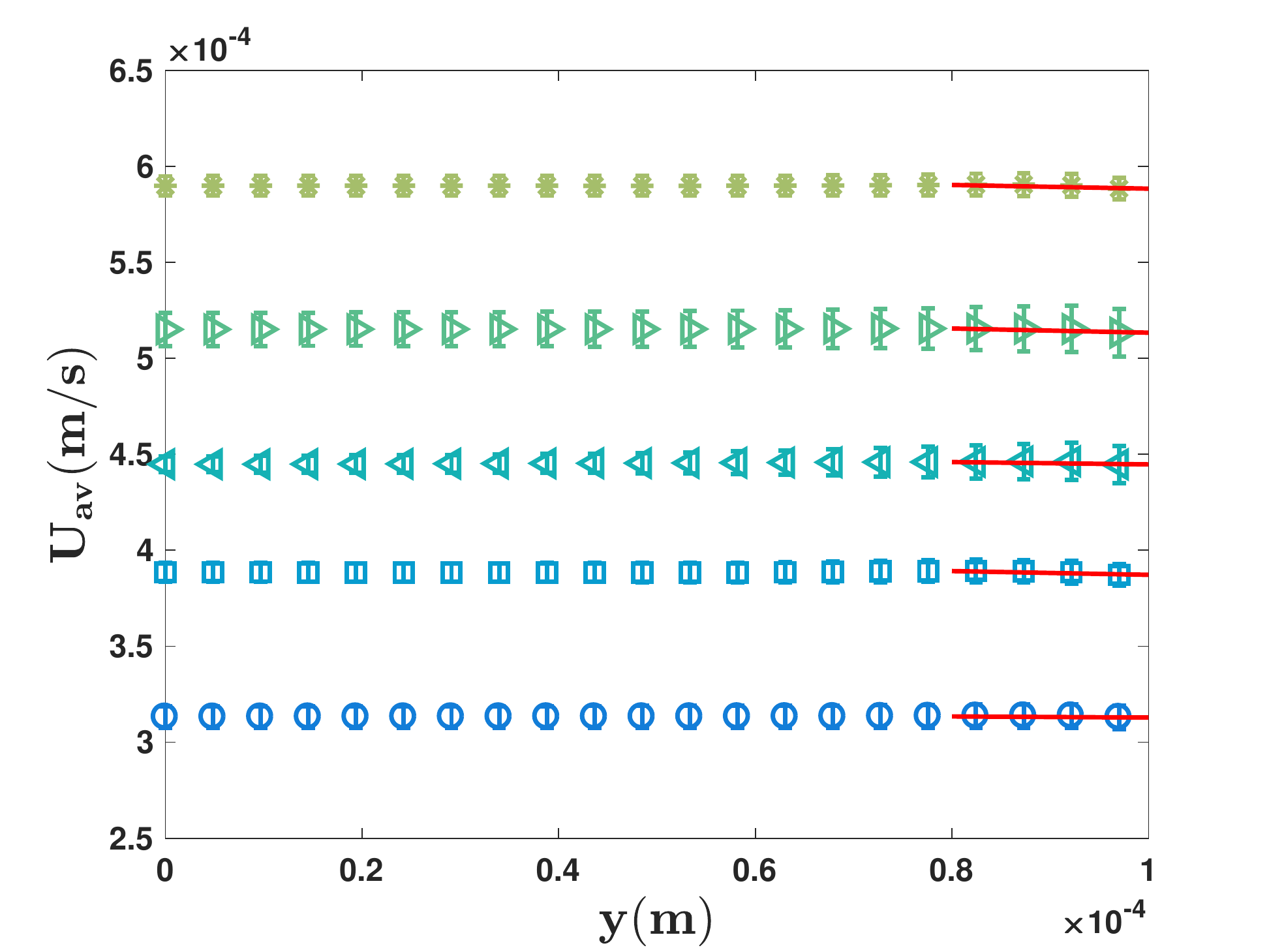}
     \label{F:Plug}
}
\subfigure[]{
                 \includegraphics [height=6cm] {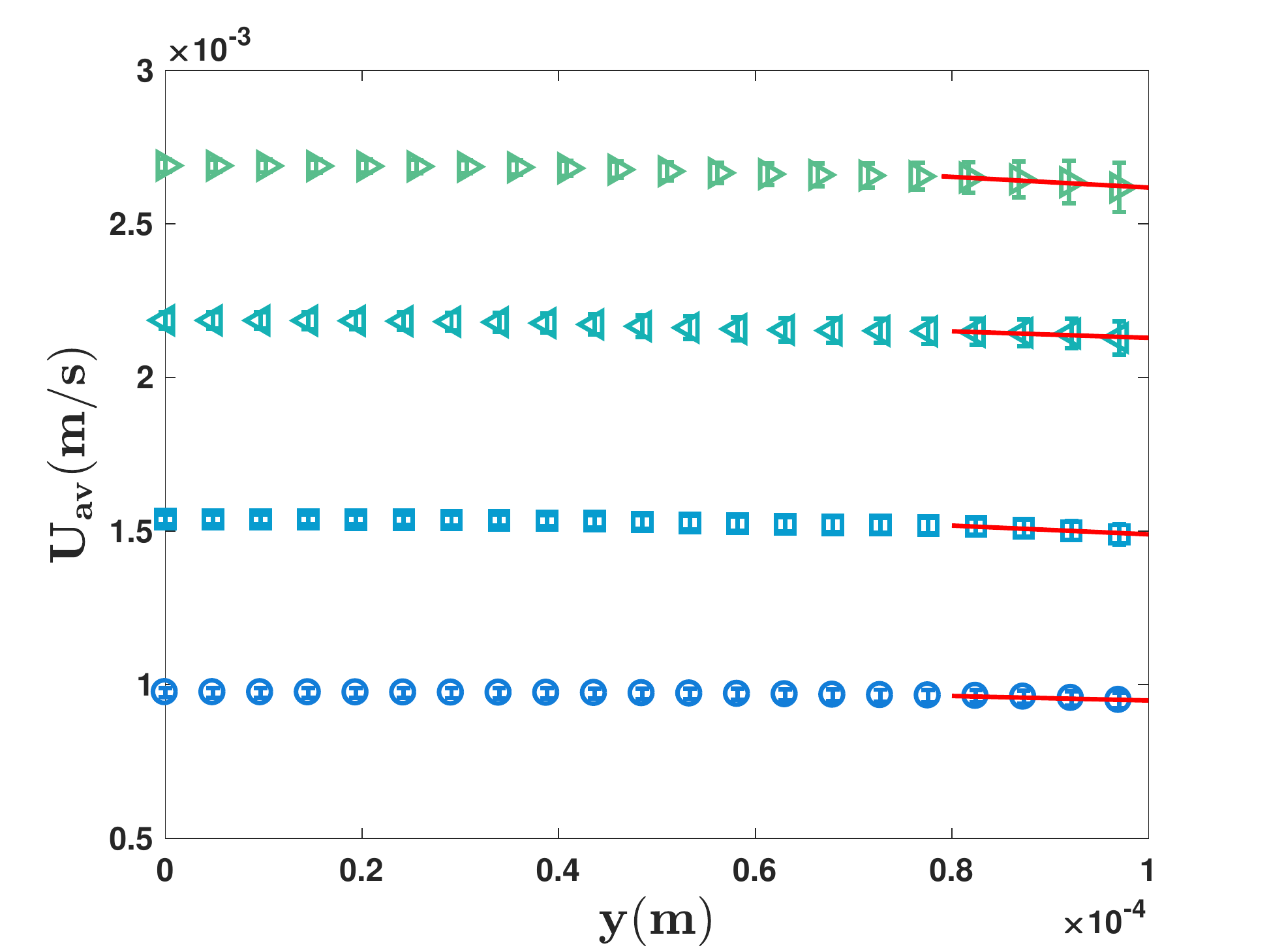}
    \label{F:Transition}
}
\subfigure[]{
                 \includegraphics [height=6cm] {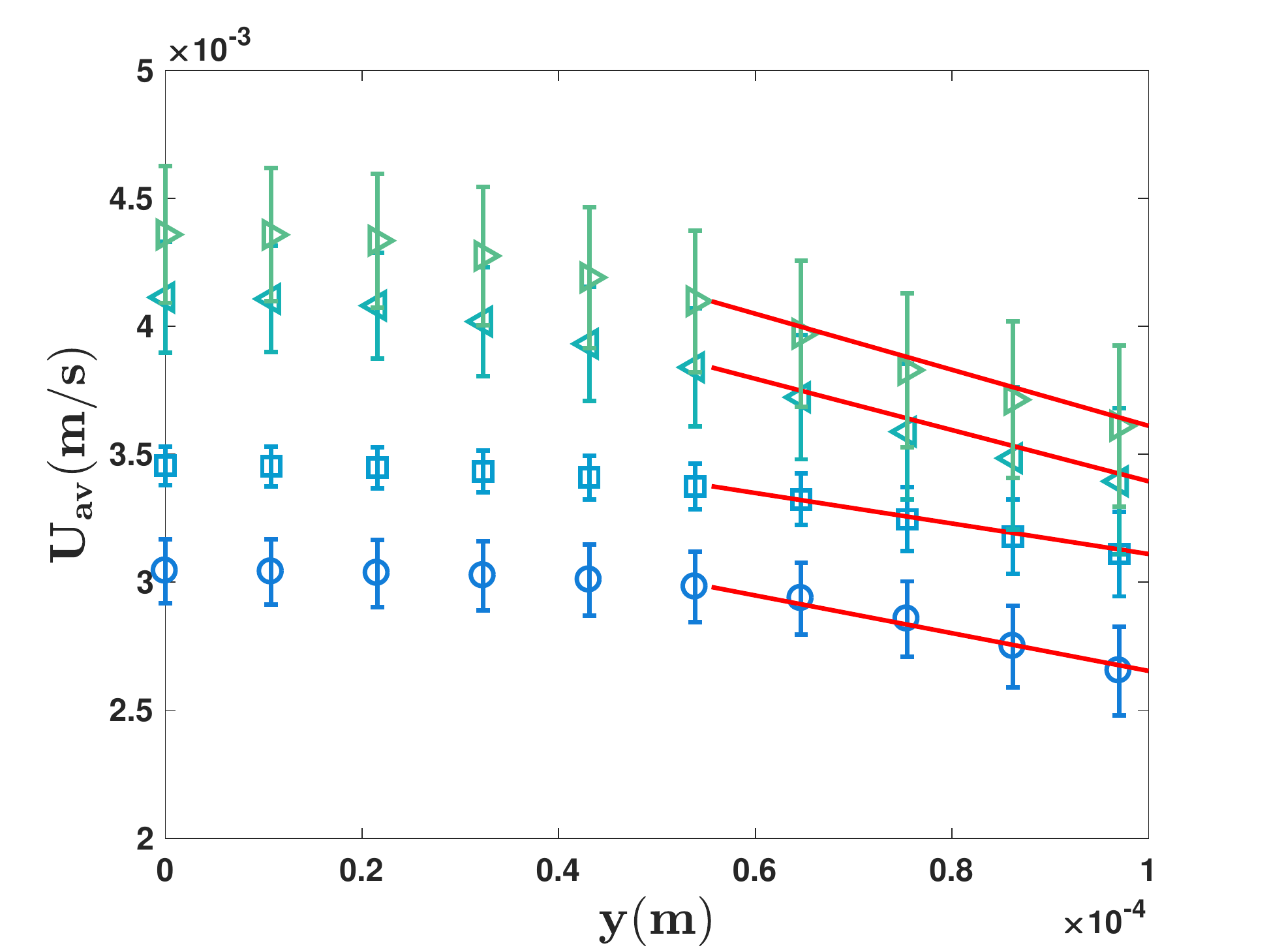}
    \label{F:fluid}
}

\caption{Velocity profiles for different values of flow rate $Q$. Panel \subref{F:Plug} corresponds to the solid regime: circles- $Q =1~\mu l/min$, squares- $Q =1.4~ \mu l/min$, left triangles- $Q =1.6~ \mu l/min$, right triangles $Q =1.8~ \mu l/min$, stars- $Q =2~ \mu l/min$. Panel \subref{F:Transition} corresponds to the  transition: circles- $Q =3.5 \mu l/min$, squares- $Q =5.5~ \mu l/min$, left triangles- $Q =8~ \mu l/min$, right triangles $Q =10~ \mu l/min$. Panel \subref{F:fluid} corresponds to the fluid regime: circles- $Q =14~ \mu l/min$, squares- $Q =15~ \mu l/min$, left triangles- $Q =21~ \mu l/min$, right triangles $Q =23~ \mu l/min$.
\label{F:velocity profiles}}
\end{figure}

Profiles of the time averaged flow velocity $U_{av}$ measured according to the method described in Sec. \ref{subsec:PIV measurements} are presented in Fig.  \ref{F:velocity profiles}. 
Due to the axial symmetry of the flow, velocity profiles were plotted only across half width of the channel ($y=0$ corresponds to the centreline of the micro-channel and $y=1\times 10^{-4}~m$ corresponds to the channel wall.).
The standard deviation of the velocity profiles was smaller than $5\%$ for different values of flow rates.  Depending on the magnitude of the constant imposed flow rate $Q$, three distinct flow regimes are observed in Fig.  \ref{F:velocity profiles}.

At low driving flow rates ($Q < 2~ \mu l/min$), a full plug flow regime is observed in Fig. \ref{F:Plug}. The entire Carbopol gel is un-yielded and \emph{``slides"} over a very thin liquid layer located in the proximity of the channel walls. The fully plug velocity profiles are characteristic of a solid-like behavior visible in Fig. \ref{F:flow_curve}, region \textbf{(S)} and phenomenologically consistent with the observations of P\'{e}rez-Gonz\'{a}lez et al. \cite{RheoPIV}.

As the flow rate is gradually increased past this first flow regime, a second flow regime is observed. The Carbopol gel is partially yielded in the proximity of the channel walls (where the velocity gradients are the largest) but a central un-yielded plug may still be observed around the centre-line of the micro-channel, Fig. \ref{F:Transition}. The shear stresses are not large enough to yield the material along the entire width of the channel. This behavior might be related to the solid-fluid transition of the material observed in Fig. \ref{F:flow_curve}, region \textbf{(S+F)}. 

As the flow rate is increased even further ($Q > 10~ \mu l/min$), the Carbopol gel is fully yielded across the entire width of the channel. The further increase of flow rate translates into additional shear stresses and a shear thinning flow regime is observed in Fig. \ref{F:fluid} (the local shear stresses exceed the yield stress everywhere across the channel width). The shear thinning velocity profiles are characteristic of a fluid-like regime visible in Fig. \ref{F:flow_curve}, region \textbf{(F)}.

Corresponding to each flow regime, the velocity profiles clearly exhibit a non-zero slip behavior at the channel walls (a non-zero slip velocity at $y=1\times 10^{-4}~ m$). The measured slip velocity is systematically larger than the error bar of the velocity profile near the wall.

\begin{figure}
\centering
{\includegraphics[height=7cm]{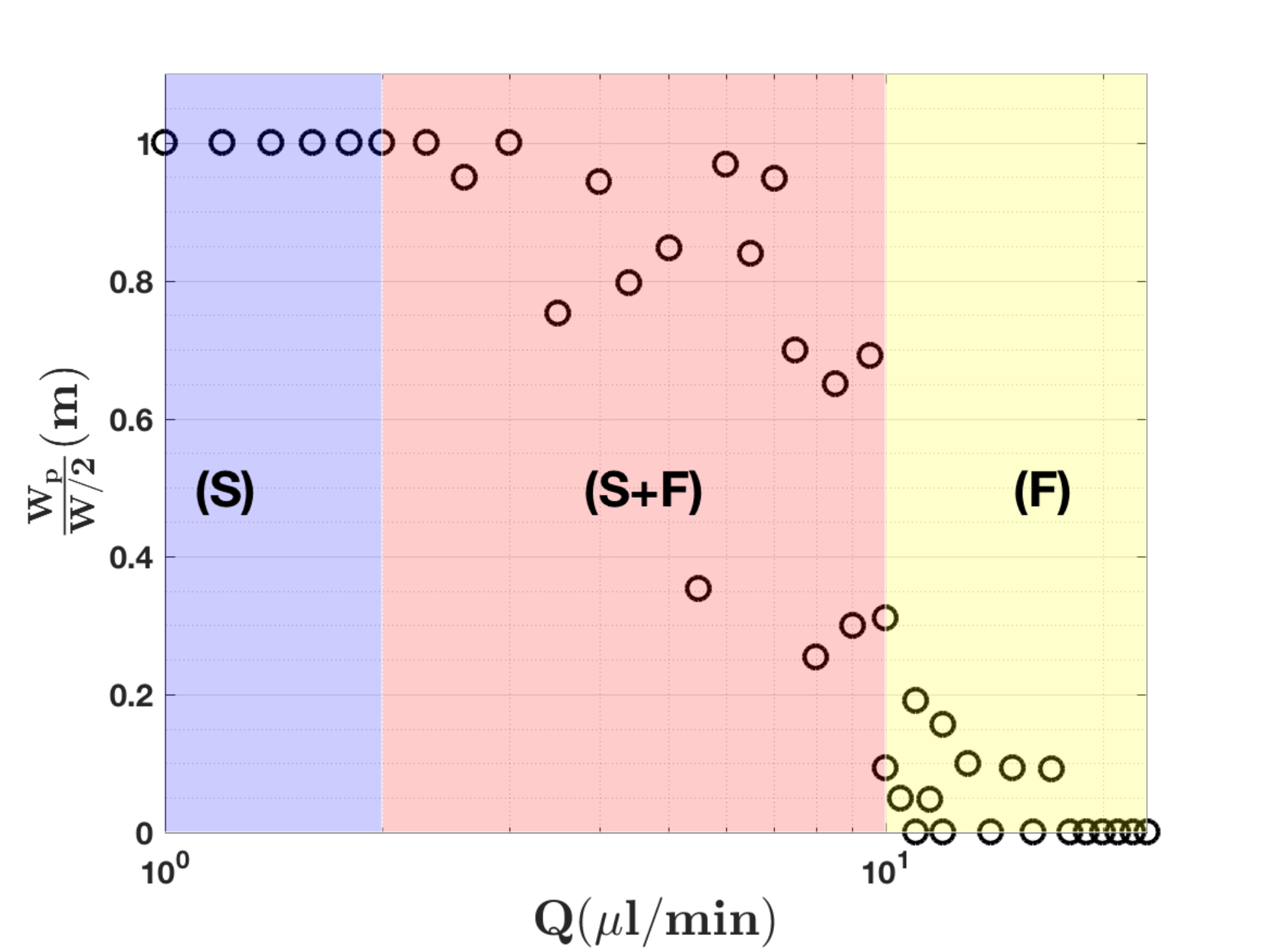}}
\caption{Dependence of the normalized plug width $2W_p/W$ on the applied flow rate $Q$. The symbols marking the highlighted regions denote the deformation regimes and are explained in the text: \textbf{(S)} - solid, \textbf{(S + F)} - solid-fluid coexistence, \textbf{(F)} - fluid.\label{F:Plug radius}}
\end{figure}

The measured velocity profiles allow one to compute the velocity gradients near the wall, the slip and plug velocities and the plug size $W_p$ defined by the extent of the flat central region. The evolution of the size of the rigid plug $W_p$ with the flow rate is presented in Fig. \ref{F:Plug radius}. At low values of flow rate ($Q <2~ \mu l/min$) the plug size is practically independent on the flow rate: the plug fills the entire width of the channel. Beyond ($Q\approx2~ \mu l/min$) the plug size decreases as $Q$ is increased. The shear stresses are not large enough to yield the material along the entire width of the micro-channel. Beyond a critical value of the flow rate  ($Q >10~ \mu l/min$), a plug  is no longer observed ($W_p \approx 0 $) and the material becomes fully yielded  (the local shear stresses exceed the yield stress everywhere across the channel width).

\subsection{Scaling behavior at the wall within various flow regimes} \label{sec:results_wallslipbehavior}
As already emphasised in the introduction, understanding the scaling behavior of the relevant hydrodynamic quantities (shear stress, shear rate, slip velocity) in the proximity of the wall is of paramount importance to developing reliable microscopic models able to explain the confined flows of viscoplastic materials in the presence of wall slip and for the implementation of numerical simulations. Based on the characterization of the flow fields presented in Sec. \ref{sec:results_flowregimes}, we study in this section the scaling behavior for each flow regime previously identified.  

\begin{figure}
\centering
\subfigure[]{
     \includegraphics [height=7cm]{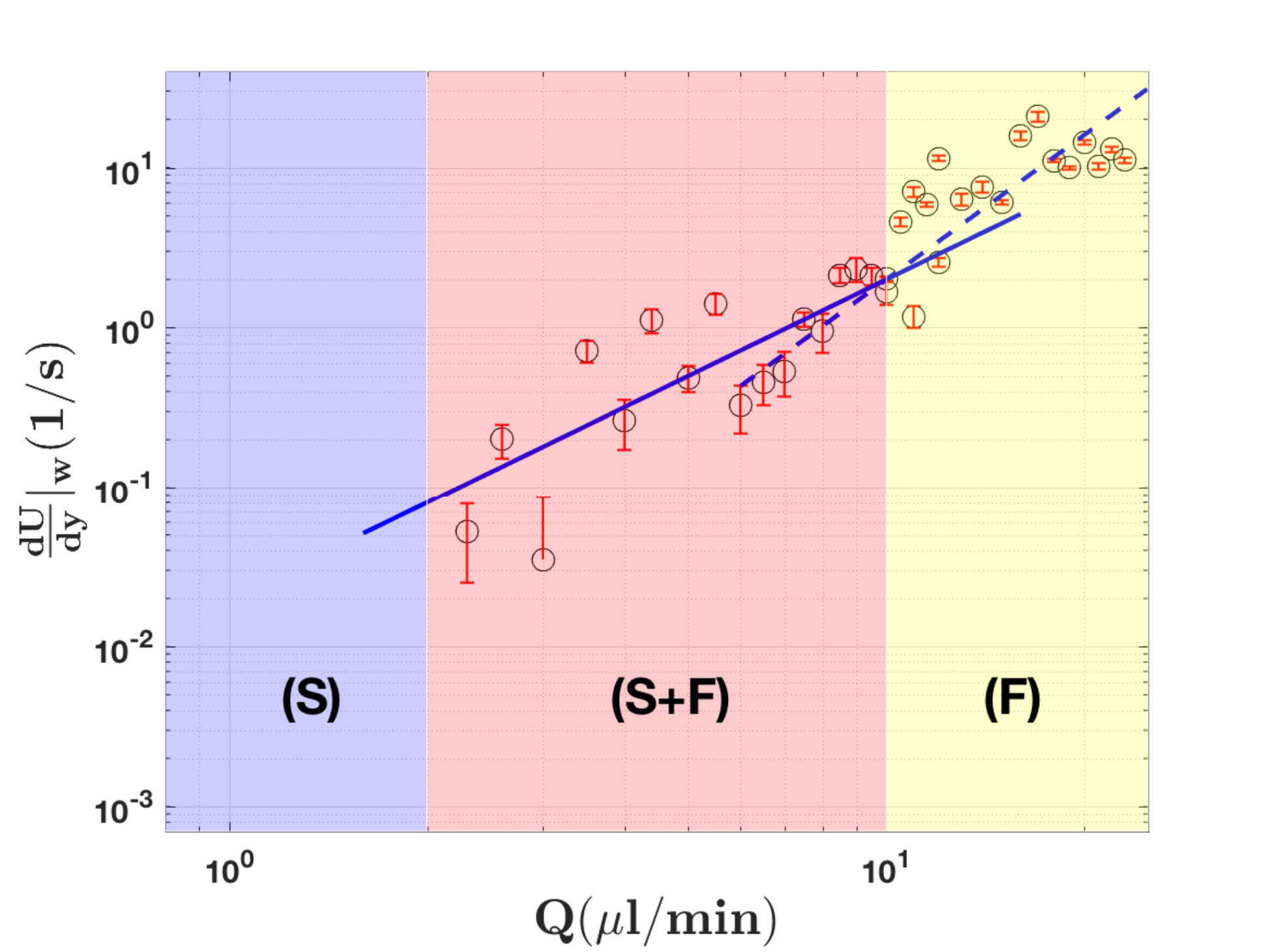}
     \label{F:dU_vs_Q}
}
\subfigure[]{
      \includegraphics [height=7cm] {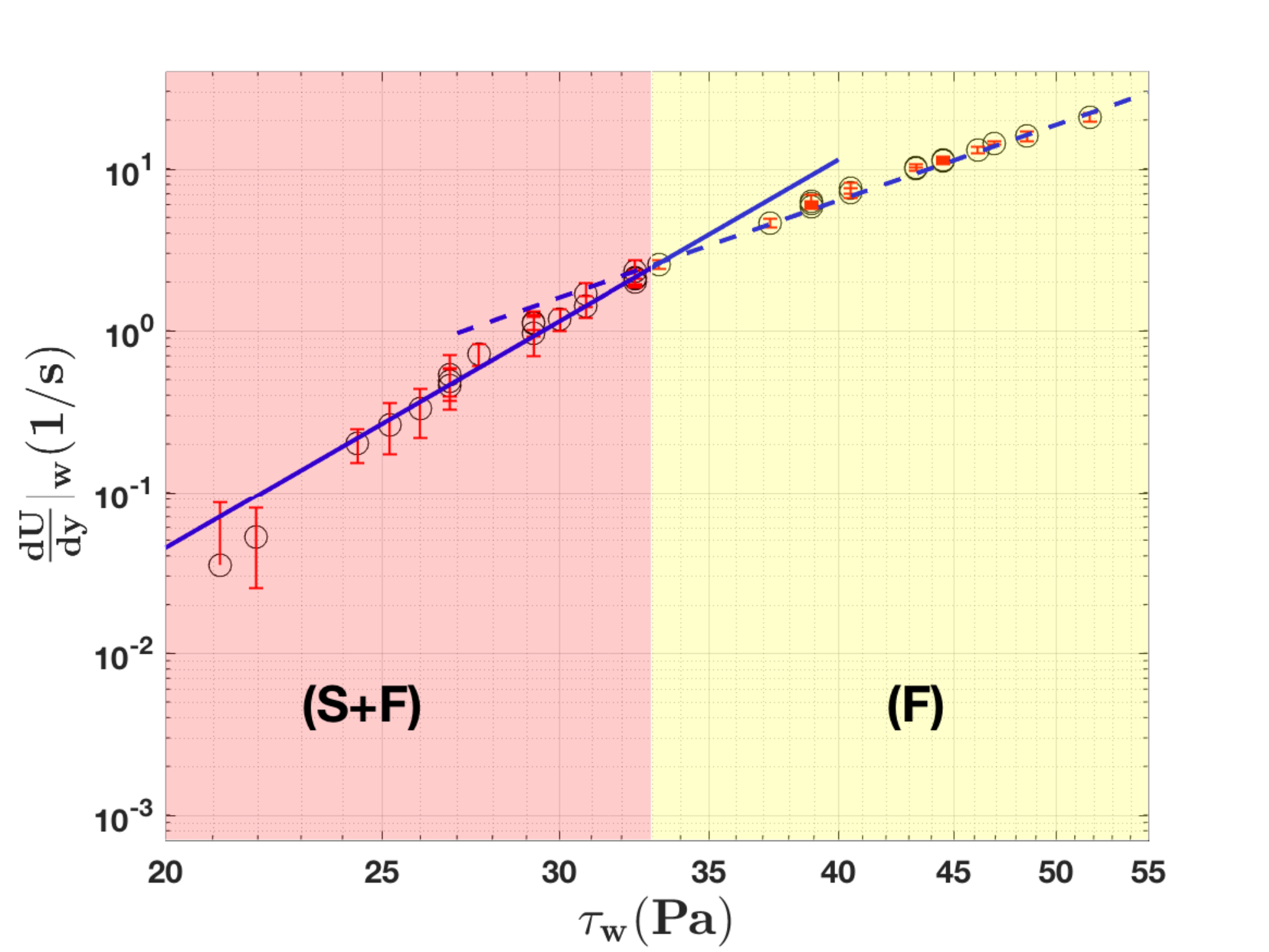}
      \label{F:dU_vs_shearstress}   
}      
 \caption{ \subref{F:dU_vs_Q} Dependence of the wall velocity gradient on the applied flow rate $Q$. The solid line (\textcolor{blue}{\textbf{-}}) and the dashed line (\textcolor{blue}{\textbf{- -}}) are guides for the eye, $\frac{dU}{dy} \vert _w\propto Q^{2}$ and $\frac{dU}{dy} \vert _w\propto Q^{3}$ respectively. \subref{F:dU_vs_shearstress} Dependence of the wall velocity gradient on the wall shear stress computed using the rheological measurements and computed wall velocity gradients. The solid line (\textcolor{blue}{\textbf{-}}) and the dashed line (\textcolor{blue}{\textbf{- -}}) are guides for the eye, $\frac{dU}{dy} \vert _w\propto \tau_w^{7.99}$ and $\frac{dU}{dy} \vert _w\propto \tau_w^{4.8}$ respectively.}  
\label{F:dU}
\end{figure}

The dependence of the wall velocity gradients computed according to the procedure detailed in Sec. \ref{subsec:PIV measurements} on the flow rate $Q$ is presented in Fig.  \ref{F:dU_vs_Q}. In the solid region  \textbf{(S)}, the material is fully unyielded and, within the instrumental accuracy of our technique,  no velocity gradients could be reliably measured near the solid wall. As the flow rate $Q$ is gradually increased, one can clearly notice the existence of two different scaling regimes. Within the solid-fluid transition regime \textbf{(S+F)}, the wall velocity gradient scales with the flow rate as $\frac{dU}{dy} \vert _w \propto Q^{2}$ (see full line in Fig. \ref{F:dU_vs_Q}). A further increase of the flow rate into the fluid regime  \textbf{(F)} leads to a different scaling of the wall velocity gradient, $\frac{dU}{dy} \vert _w \propto Q^{3}$ (see dashed line in Fig. \ref{F:dU_vs_Q}).

%
The measurements of the wall velocity gradients $\frac{dU}{dy} \vert _w$ allow one to compute the wall shear stresses $\tau_w$ using the macro-rheological measurements presented in Fig. \ref{F:flow_curve}. To do so, we interpreted $\frac{dU}{dy} \vert _w$ as the relevant scale for the rate of shear and interpolated the corresponding stress from the macro-rheological flow curve presented in Fig. \ref{F:flow_curve}.
The result of this evaluation is presented in Fig. \ref{F:dU_vs_shearstress}. Two different scaling regimes of the wall shear rate with the wall shear stress can  be clearly observed. The first scaling regime corresponds to a partial yielding of the gel in the vicinity of the wall, $\frac{dU}{dy} \vert _w\propto \tau_w^{7.99}$ (see solid line in Fig. \ref{F:dU_vs_shearstress}). Above a critical value of wall shear stress and corresponding to the fluid deformation regime a second scaling regime is observed,  $\frac{dU}{dy} \vert _w \propto \tau_w^{4.8}$ (see dashed line in Fig. \ref{F:dU_vs_shearstress}). 

The scaling exponent obtained by fitting the data corresponding to the fully yielded regime differs from the power law exponent obtained by the Herschel-Bullkley fitting to the macro-rheological measurements. This difference might be attributed to the confinement of the flow as suggested in Refs. \cite{Geraud2013, LIU201825}.
However, we can notice that the value of the critical wall shear stress defined as the intersection between the two power laws is close to the yield stress measured in the rheometric flow and fitted by the Herschel-Bulkley model, see Fig. \ref{F:flow_curve}.  

\begin{figure}
\centering
\subfigure[]{
     \includegraphics [height=7cm] {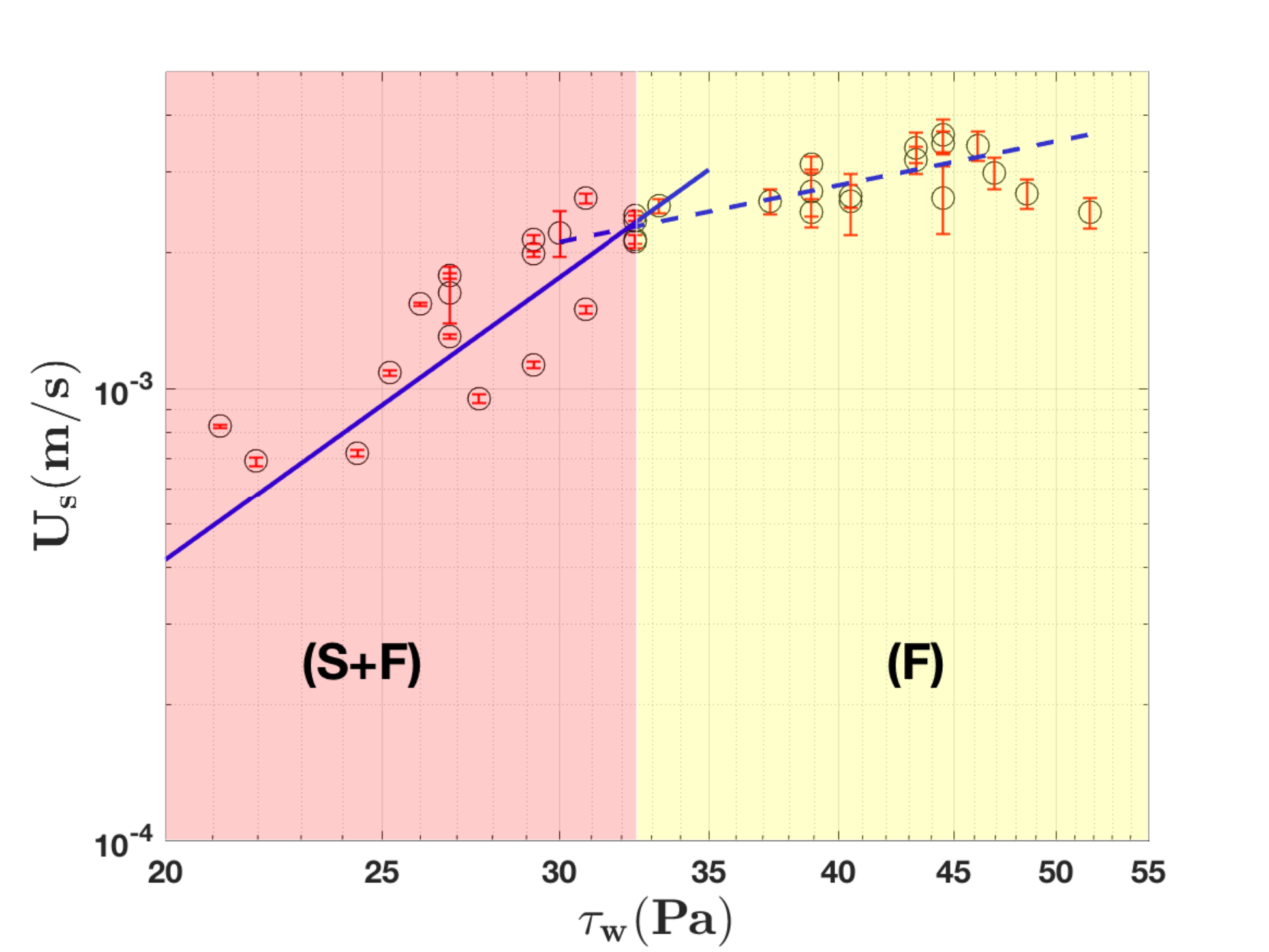}
     \label{F:Us_vs_shear}
}
\subfigure[]{
                 \includegraphics [height=7cm] {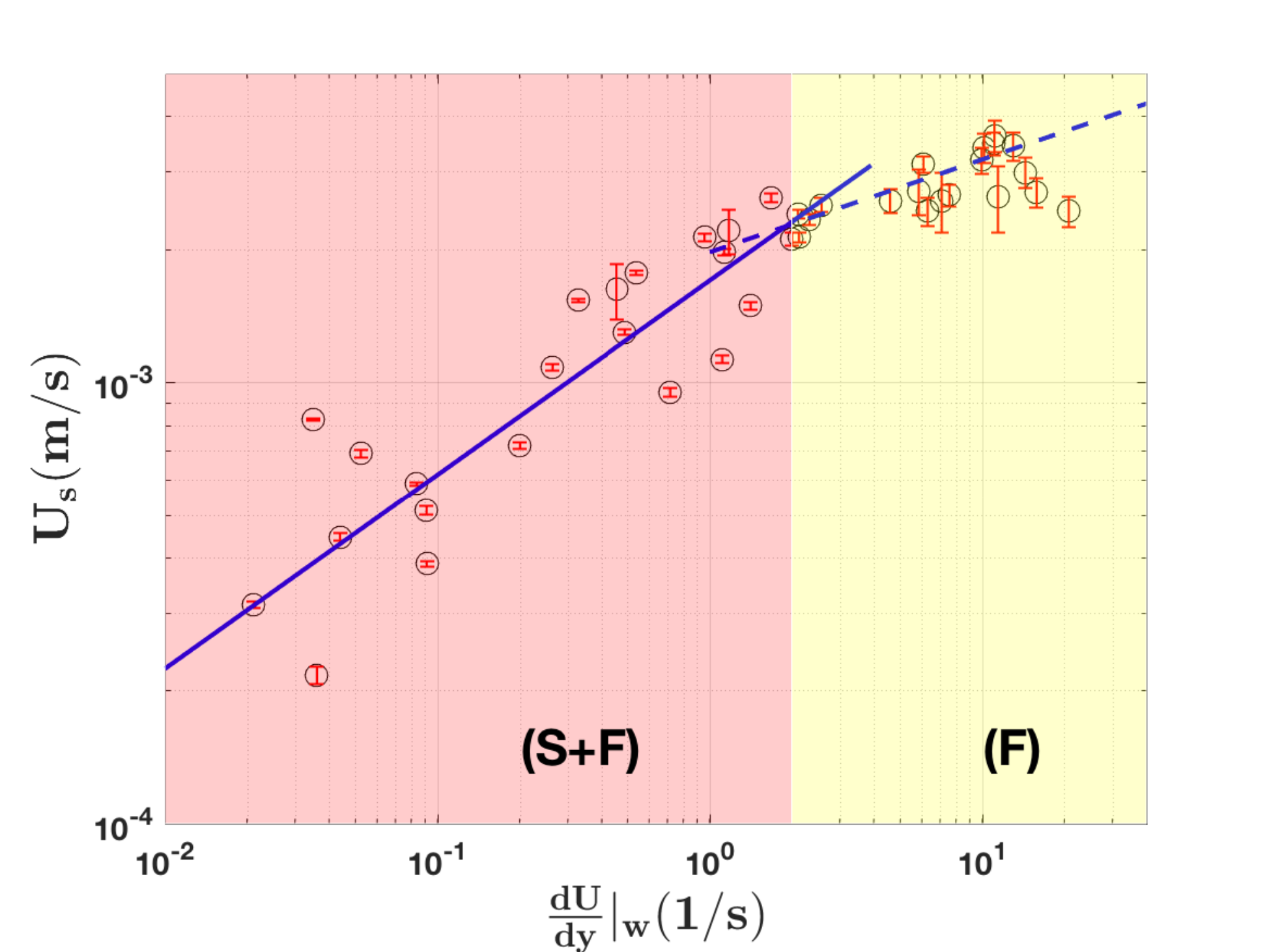}
    \label{F:Us_vs_dU}
}

\caption{\subref{F:Us_vs_shear} Dependence of the wall slip velocity $U_s$ on the wall shear stress $\tau_w$. The full (\textcolor{blue}{\textbf{--}}) and dashed (\textcolor{blue}{\textbf{-- --}}) lines  are guides for the eye, $Us\propto \tau_w^{3.55}$, $Us\propto \tau_w$ respectively. 
\subref{F:Us_vs_dU}  Dependence of the wall slip velocity $U_s$ on the wall velocity gradient $\frac{dU}{dy} \vert _w$. The full (\textcolor{blue}{\textbf{--}}) and dashed (\textcolor{blue}{\textbf{-- --}}) lines  are guides for the eye, $Us\propto \left(\frac{dU}{dy} \vert _w\right)^{0.44}$ and $Us\propto \left(\frac{dU}{dy} \vert _w \right)^{0.12}$ respectively. The symbols marking the highlighted regions denote the deformation regimes and are explained in the text: \textbf{(S + F)} - solid-fluid coexistence, \textbf{(F)} - fluid.}
\label{F:Us}
\end{figure}

The dependence of the slip velocity $U_s$ on the wall shear stress $\tau_w$ is presented in Fig. \ref{F:Us_vs_shear}. Because in the solid regime the wall velocity gradients are too small to be reliably measured, we could not use the method described above to reliably compute the wall shear  stresses. For higher flow rates (beyond the \textbf{(S)} regime) , again, two distinct scaling laws are observed. Corresponding to the solid-fluid coexistence regime \textbf{(S+F)}, $U_s \propto \tau_w^{3.55}$. Within the fully yielded regime \textbf{(F)}, $U_s \propto \tau_w$. 

The dependence of the slip velocity $U_s$ on the wall velocity gradients $\frac{dU}{dy} \vert _w$  is illustrated in Fig. \ref{F:Us_vs_dU}.  Within the \textbf{(S + F)} and \textbf{(F)}  $\frac{dU}{dy} \vert _w$ scales as $Us\propto \left(\frac{dU}{dy} \vert _w\right)^{0.44}$ and $Us\propto \left(\frac{dU}{dy} \vert _w \right)^{0.21}$, respectively.

\subsection{Comparison with previous results from the literature} \label{sec:comparison}
We compare in the following the scaling laws described in Sec. \ref{sec:results} with results obtained by others.

 P\'{e}rez-Gonz\'{a}lez et al. have investigated the scaling behaviour in a millimetre size glass channel, \cite{RheoPIV}. They  have estimated the apparent shear rate $\dot \gamma_a$ and the wall shear stress $\tau_w$ as $\dot\gamma_a=\frac{32 Q}{\pi D^3}$, $\tau_w=\frac{\Delta P}{4 \frac{L}{D}}$ respectively, where $\Delta P$ is the pressure drop between the capillary ends, $L$ and $D$ are respectively the length and the diameter of the borosilicate glass capillary and $Q$ is the flow rate calculated from the integration of the velocity profiles. Consistently with our findings, they identify three distinct flow regimes: a purely plug flow before yielding, solid- liquid transition and shear thinning at relatively high shear rates. The apparent shear rate $\dot \gamma_a$ scales with the wall shear stress $\tau_w$ as  $\dot \gamma_a\propto\tau_w^{2.44}$ in the fluid regime. This scaling law differs significantly from the one we observe, $\frac{dU}{dy} \vert _w\propto \tau_w^{4.8}$ (the dashed line in Fig. \ref{F:dU_vs_shearstress}). 
This difference might be related to the way $\dot \gamma_a$ is calculated ($\dot\gamma_a=\frac{32 Q}{\pi D^3}$). This formula assumes that the velocity profile is parabolic and there exists no velocity slip at the walls which, obviously, is not the case for a Carbopol gel flowing in a glass capillary.

By direct observation of a rheometric flow of a microgel, Meeker et al. in Ref. \cite{cloitreslip} observe that $U_s$ remains constant for $1< \frac{\tau}{\tau_y}<1.5$ while in our experiments the slip velocity increases linearly with the wall shear stress, $Us\propto \tau_w$. 

The same authors observe in a subsequent publication three slip regimes in a macro-rheological setup, \cite{R2}. In a solid regime they find a slip velocity that scales as the stress squared whereas in the intermediate deformation regime they find a constant slip velocity. These scaling results are obtained within the framework of a phenomenological microelastohydrodynamic lubrication. As we use no model for assessing the scaling behaviour, a comparison is rather difficult to make.
 
 P\'{e}m\'{e}ja and coworkers have obtained two wall slip scaling laws for the micro-channel flow of a Carbopol gel: for low wall shear stresses $\tau_w<100~Pa$ $U_s \propto \tau_w ^2$ whereas for higher stresses they obtain a linear scaling. The quadratic scaling law is at odds with the result presented in Fig. \ref{F:Us_vs_shear}, $U_s. \propto \tau_w^{3.55}$ whereas the linear scaling law is consistent with our findings. This partial agreement may be explained by the way  P\'{e}m\'{e}ja compute the wall shear stress using the driving pressure drop $\Delta p$, $\tau_w = \frac{h}{2L} \Delta p$ ($L$ being the length of their micro-channel and $h$ its depth). This correlation may become relatively accurate \emph{"only"} far from the yield point whereas it is most probably formal in the solid \textbf{(S)} and in the intermediate \textbf{(S+F)} regimes.
 
To describe the wall slip behavior Kaylon proposed the following relationship between the slip velocity and the wall shear stress \cite{kaylonslip}:
\begin{equation}\label{Kaylon}
U_s  =\beta (\tau_w)^{1/n_b}
\end{equation}
Here $\beta$ relates the slip layer $\delta$ to the consistency of the binder fluid $m_b$ and to its power-law index $n_b$, $\beta=\frac{\delta}{m_b^{1/n_b}}$. This simple phenomenological scaling relationship is derived based on the assumption of the existence of a depleted layer of solvent of width $\delta$ in the proximity of the solid boundary (such assumption is often used in order to derive simple phenomenological scaling models, e.g. see Ref. \cite{R9}).
The Carbopol gel is considered as jammed system of swollen gel micro-particles \cite{piaucarbopol}. As the solvent of the Carbopol gel was deionised water, the \emph{"binder fluid"} might be considered as Newtonian,  $K_b=10^{-3} ~ Pa.s$ and $n_b=1$. 

In the fluid regime, $U_s$ scales linearly with $\tau_w$, $U_s \propto \tau_w$ (the dashed line in Fig. \ref{F:Us_vs_shear}). This linear scaling law confirms the assumption that the \emph{"binder fluid'"} is Newtonian, $n_b=1$. A similar scaling law was found by Poumaere et al. \cite{unsteady} while it differs from the ones predicted theoretically by Piau \cite{piaucarbopol} $U_s \propto \tau_w^{1/3}$ for loosely packed system and $U_s \propto \tau_w^{2}$ for closely packed system. This difference might be related to the fact that the lubricating fluid is considered as non-Newtonian fluid in Ref. \cite{piaucarbopol}.

In our study, the measured factor $\beta=70\cdot10^{-6} ~ m.Pa^{-1}.s^{-1}$ leads to a slip layer of thickness $\delta \approx 0.07~ \mu m$. This value is of the same order of magnitude as the one found by Jiang et al.  \cite{metznerslip}, $\delta \approx 0.1~ \mu m$ and by Poumaere et al. \cite{unsteady}, $\delta \approx 0.23~ \mu m$. However, this width obtained by such simple estimates is far too small to be accessible via direct visualisation of the flow meaning that, the main hypothesis used to predict this scaling remains for now just a hypothesis. This once more justifies our approach of relying on no assumption but computing the stresses by relating the steady state rheological behaviour to the in-situ measurements of the wall velocity gradients.
 
\section{On the possibility of inhibiting the wall slip in an acrylic made micro-channel} \label{sec:inhibitting_wallslip}
There exist a number of situations when the emergence of the wall slip phenomenon may turn quite problematic. The best known example relates, perhaps, to the macro-rheological characterization of pasty materials where the presence of walls leads to systematic errors particularly around the yield point. From a theoretical and numerical perspective the presence of wall slip is equally troublesome as, to our best knowledge, there exists no fundamental theory that rigorously captures the scaling behavior of the slip velocity at the wall and the choice of a slip law is often arbitrary. A direct way of eliminating the wall slip which is commonly used during macro-rheological tests performed with pasty materials is to use mechanically rough solid surfaces. Though effective in eliminating the wall slip, this solution comes at a price difficult to cope with during experiments that require the in-situ visualization of the flow structure: the loss of the optical transparency. An alternative method relates to \emph{``engineering''} a special chemical treatment that facilitates the adhesion of the microstructural elements of the pasty material to a solid wall without altering its optical transparency. For the case of a Carbopol gel, Metivier and co-workers proposed a chemical reaction able to eliminate the slip along a smooth acrylic made wall, \cite{StickSlip}. The authors of this study prove the effectiveness of their method via macro-rheological tests performed with various gaps and indicate and equally emphasise its temporal stability during $46$ days. A detailed investigation of the physical mechanisms, responsible for the apparent macroscopic effectiveness of this solution is, however, missing. After providing a detailed description of the chemical reaction, we focus in the following on a detailed analysis of this solution via both macro-rheological tests and microfluidic experiments. 
 
\subsection{Chemical treatment of the micro-channel}\label{sec:chemical}

The solution proposed by Metivier and her coworkers consists of depositing a thin layer of polyethylenimine $PEI$ (Aldrich) on the $Polymethyl$ methacrylate $PMMA$ surface.
Low concentration of $PEI$ solutions $C= 10^{-8}~wt\%-2~wt\%$ at neutral pH have been prepared by dissolving the right amount of $PEI$ in deionized water. The $PMMA$ substrate was immersed into the $PEI$ polymer solution during times ranging in between  $60~s$ and  $24~h$ and concluded that the final results is unchanged. Finally they have soaked the chemically treated substrate into double deionized water for $6~h$ in order to remove the excess (unreacted) of $PEI$. 

The idea of the chemical treatment proposed by Metivier and her co-workers actually originates from a previous study by Kitagawa et al.,  \cite{Kitagawa}. Though apparently similar, the exact details of the procedure of Kitagawa and al. are actually quite different.  
The $PMMA$ microchip was cleaned by damping it in methanol for $10~mins$ then rinsed in deionized water. A significantly  higher concentration of $PEI$ solution $C= 10~wt\% - 20~wt\%$  was used. To neutralize the $PEI$ solutions, a borate buffer was equally added to the mixture. The neutralized solution was continuously circulated through the micro-channel for $2~h$. The micro-channel was rinsed  in deionized water for another $20~mins$.

As only Metivier and her coworkers have studied the influence of the chemical treatment on the wall slip of Carbopol in Ref. \cite{StickSlip}, we have decided to follow their procedure in our study rather than that of Kitagawa et al.
We used PEI solutions of $0.5~wt\%$ concentration. We found that the $pH$ is not neutral even for low concentration of $PEI$, $pH \approx 10.7$. The immersing time of our $PMMA$ substrates was for $24~h$.

\begin{figure*}
\centering
{\includegraphics[height=3.5cm]{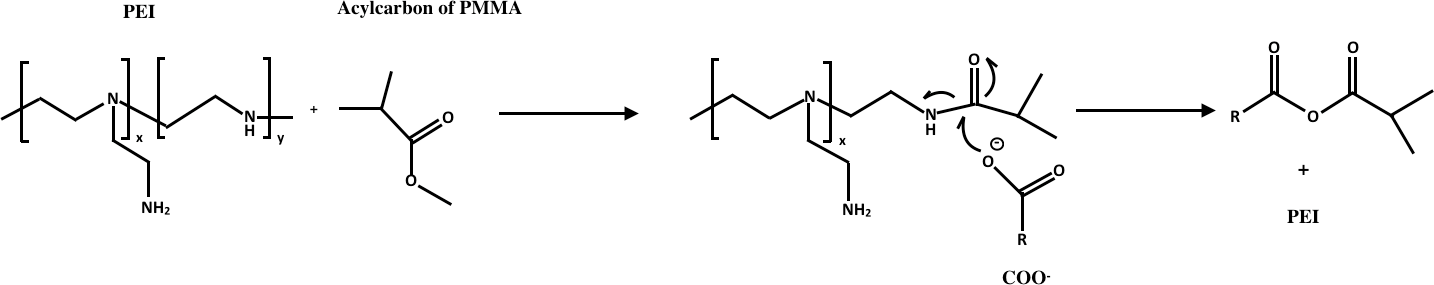}}
\caption{Chemical reaction leading to the attachement of the Carbopol gel on the PMMA surface.\label{F:Chemical_reaction}}
\end{figure*}

According to Refs. \cite{StickSlip, Kitagawa}, the idea of these methods relies on a chemical reaction between the acylcarbon groups (ester groups) of the $PMMA$ substrate with the secondary amino groups of $PEI$ and a reaction between the $COO^{-}$ groups of the Carbopol gel with the amino groups of PEI \cite{StickSlip}. After a closer analysis of the chemical reaction, we find that 
the $COO^{-}$ groups of the Carbopol gel actually bind to the amido groups as shown in Fig. \ref{F:Chemical_reaction}.

These bonds lead to the  attachment of Carbopol microgel particles to the $PMMA$ substrate and the formation of a Carbopol layer essential in preventing the wall slip. We also note that, according to Fig. \ref{F:Chemical_reaction} some $PEI$ molecules remain in the solution. 

\subsection{Assessment of wall slip behavior via macro-rheological measurements}
\begin{figure}
\centering
{\includegraphics[height=6cm]{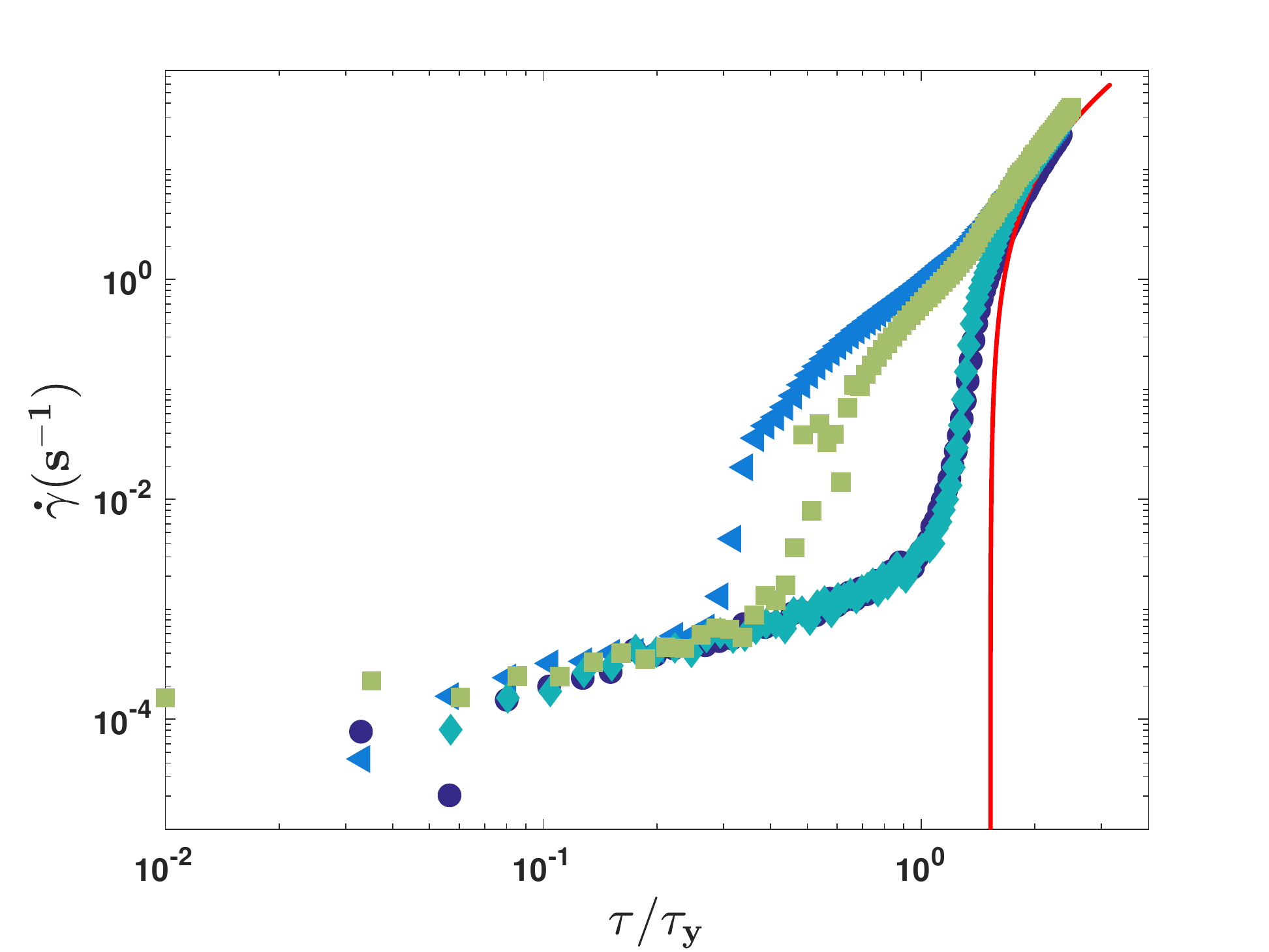}}
\caption{Macro-rheological measurements of the dependence of the rate of deformation $\dot \gamma$ on the reduced applied stress for various slip conditions on the bottom plate:  circles - no slip (glass paper glued on the bottom plate), rhomb - chemically treated acrylic block used as a bottom plate immediately after the chemical treatment was completed,  squares - chemically treated acrylic block used as a bottom plate one week after the chemical treatment,  triangles -untreated acrylic block used as a bottom plate. The full line is a fit according to the Herschel-Bulkley model. \label{F:slip_effect}}
\end{figure}

First, we have assessed the effectiveness of the chemical treatment by means of macro-rheological tests. For this purpose, we have machined two identical acrylic blocks and treated only one of them according to the protocol described in Sec. \ref{sec:chemical} while leaving the second chemically unaltered. The acrylic blocks have been used as bottom plates for macro-rheological tests. We note that the macro-rheological experiments with the chemically treated acrylic block were performed immediately after the treatment was completed thus avoiding to deal with the temporal stability of the chemical treatment.  
For all the experiments we have performed, glass paper was glued on the top plate (as described in Sec. \ref{subsec:Rheological measurments}). Thus, if any, the wall slip might solely occur on the bottom plate. The results of this comparative macro-rheological investigation are summarised in Fig. \ref{F:slip_effect} which presents the dependence of the measured rate of deformation $\dot \gamma$ on the reduced applied stress $\tau/\tau_y$ for several slip conditions on the bottom plate.
When a chemically treated acrylic block is used as a bottom plate (the stars in Fig. \ref{F:slip_effect}), the macro-rheological measurements are practically identical to those performed in the absence of wall slip (when both plates are covered with glass paper, the circles in Fig. \ref{F:slip_effect}). On the other hand, the measurements performed with the untreated acrylic block as a bottom plate (the triangles in Fig. \ref{F:slip_effect}) deviate strongly from the measurements performed in the absence of slip particularly around the yield point, $\frac{\tau}{\tau_y} \approx 1$.   
To conclude this part, from a macro-rheological perspective, the wall slip effects are negligible when the acrylic surfaces of the rheometric geometry are chemically treated and the rheological tests are conducted immediately after the treatment.  

This conclusion is in a general agreement with the main point made in Ref. \cite{StickSlip}. To this end, however, the microscopic physical mechanism behind the apparent macroscopic effectiveness of this chemical treatment remains elusive. To gain further insights into this, we focus in the upcoming section on a detailed characterization of flows in chemically treated micro-channels and on their comparison with the flows in the presence of wall slip described in Sec. \ref{sec:results}.

\subsection{Slip behavior in chemically treated micro-channels} \label{sec:treated_micro}

To understand the impact of the chemical treatment on the microscopic flows we have performed measurements similar to those performed in the presence of wall slip with chemically treated micro-channels. The operating flow rates were chosen in the same range as in the case of the experiments described in \ref{sec:results}.

\begin{figure}
\centering
\subfigure[]{
                 \includegraphics [height=6cm] {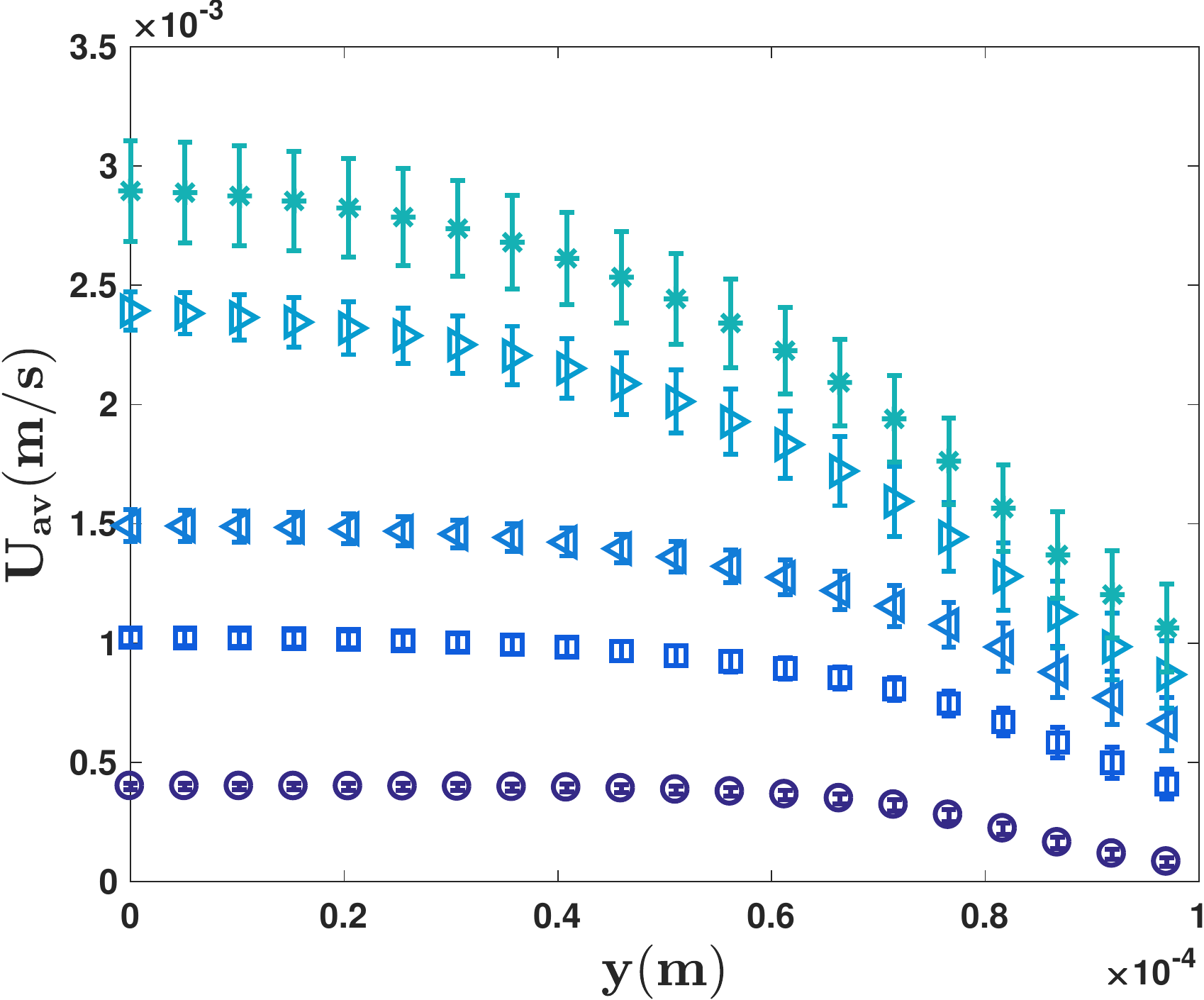}
    \label{F:Fluid_noslip}
}
\subfigure[]{
     \includegraphics [height=6cm] {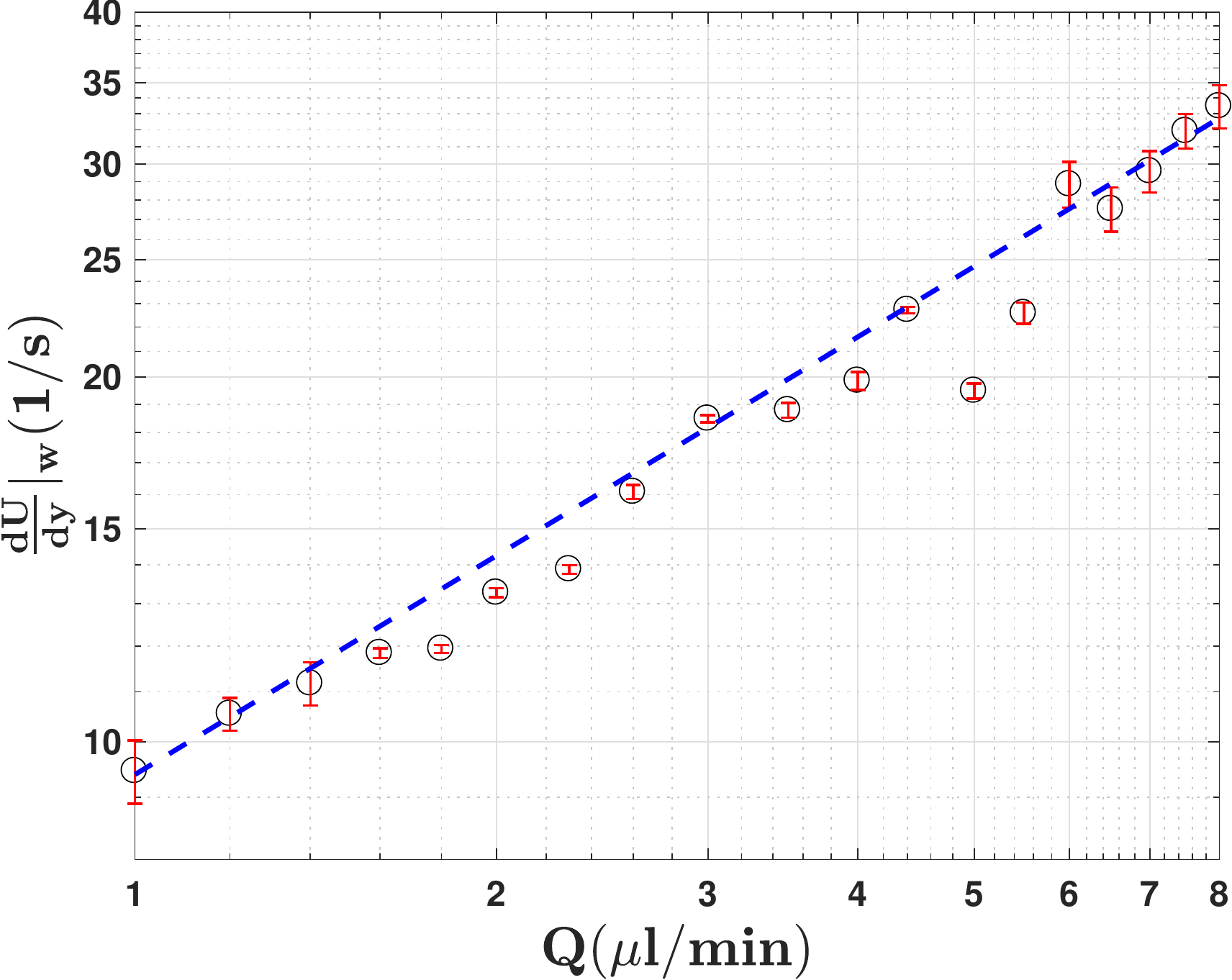}
     \label{F:dU_vs_Q_noslip}
}

\caption{\subref{F:Fluid_noslip}  Velocity profiles measured in the chemically treated micro-channel at various flow rates: circles- $Q =1 \mu l/min$, squares- $Q =2.6~ \mu l/min$, left triangles- $Q =4~ \mu l/min$, right triangles- $Q =6.5~ \mu l/min$, stars- $Q =8~ \mu l/min$. \subref{F:dU_vs_Q_noslip} Dependence of the wall velocity gradients measured in the chemically treated micro-channel on the applied flow rate $Q$. The dashed line (\textcolor{blue}{\textbf{-- --}}) is a guide for the eye, $\frac{dU}{dy} \vert _w\propto Q^{0.6}$.}
\label{F:slope_velocity_treatment}
\end{figure}

 Velocity profiles measured in the chemically treated micro-channel for several flow rates are presented in fig. \ref{F:Fluid_noslip}.  Unlike in the case of the untreated channels, we no longer observe neither a full plug flow regime as illustrated in Fig. \ref{F:Plug} nor an intermediate regime as in Fig. \ref{F:Transition}.  For each flow rate $Q$, the velocity profiles are qualitatively similar to those observed in the yielded regime illustrated in Fig. \ref{F:fluid}.
 
 A direct consequence of the chemical treatment relates to larger velocity gradients in the proximity of the wall as well as to smaller slip velocities as visible in Fig. \ref{F:dU_vs_Q_noslip}. 
 However, these measurements clearly indicate the slip at the wall is not completely removed by the chemical treatment but only inhibited. Thus, at a microscopic scale, the effectiveness of the chemical treatment is only partial. 
  
\begin{figure}
\centering
\subfigure[]{
     \includegraphics [height=5cm] {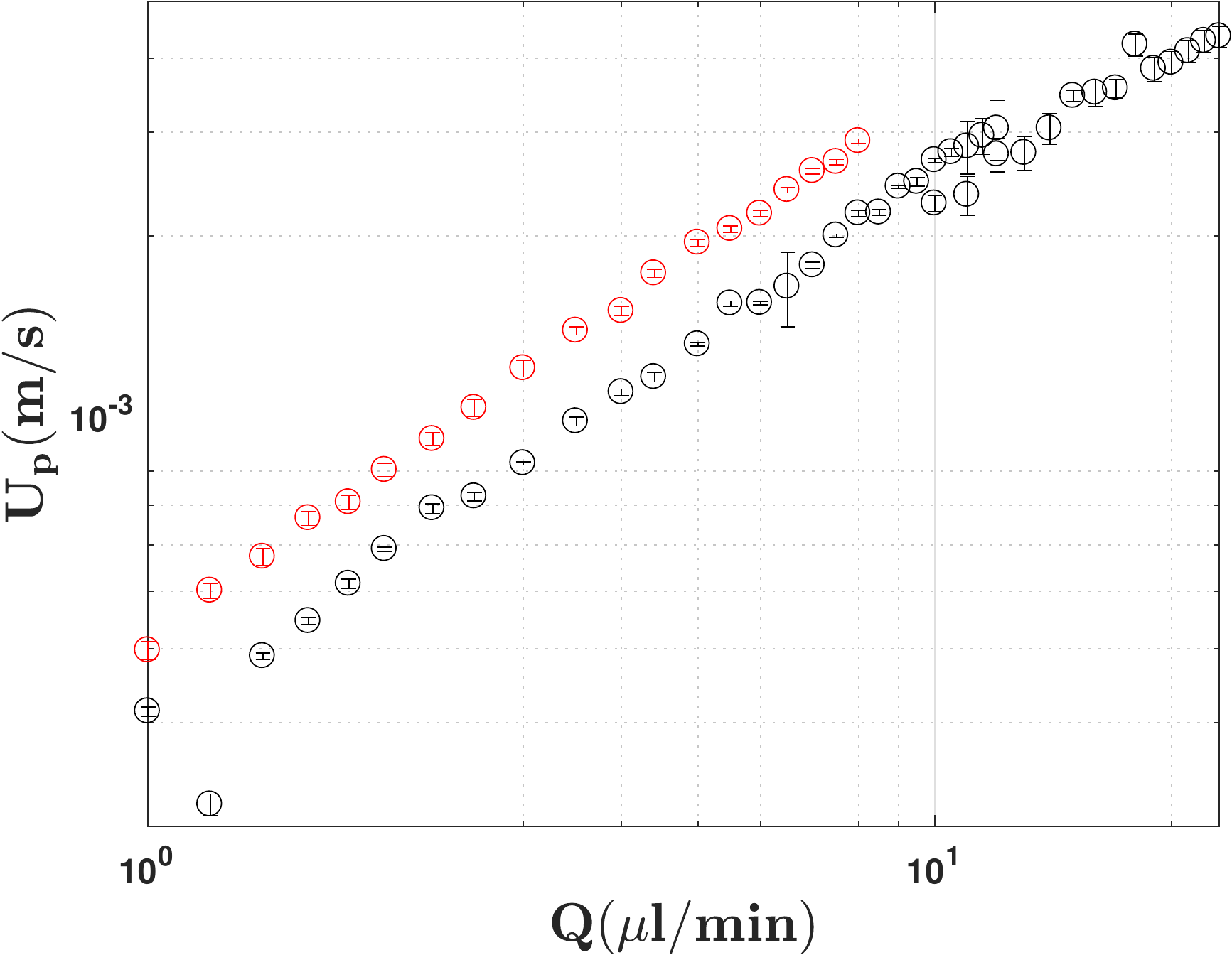}
     \label{Up_Q}
}
\subfigure[]{
                 \includegraphics [height=5cm] {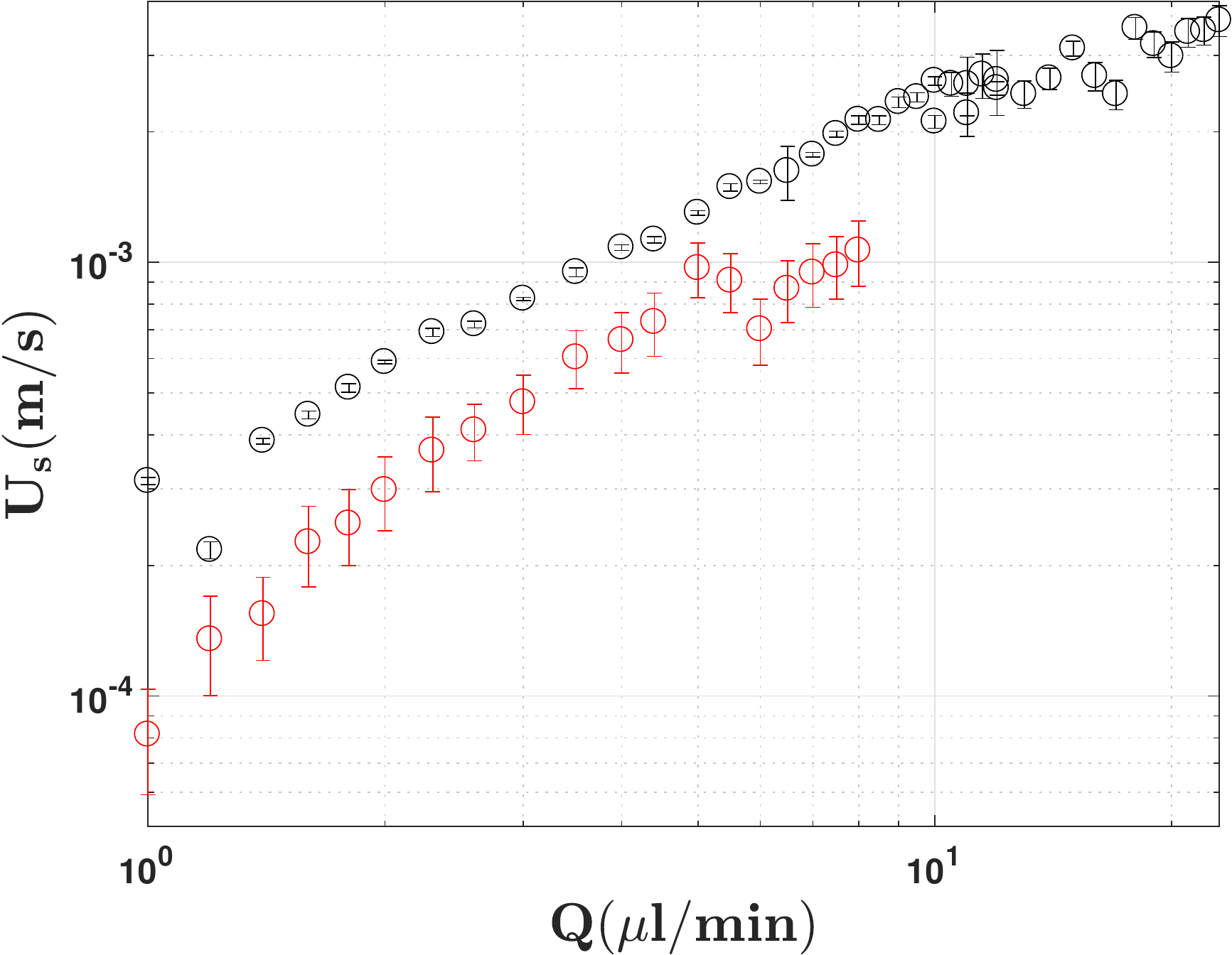}
    \label{Us_Q}
}
\subfigure[]{
                 \includegraphics[height=5cm] {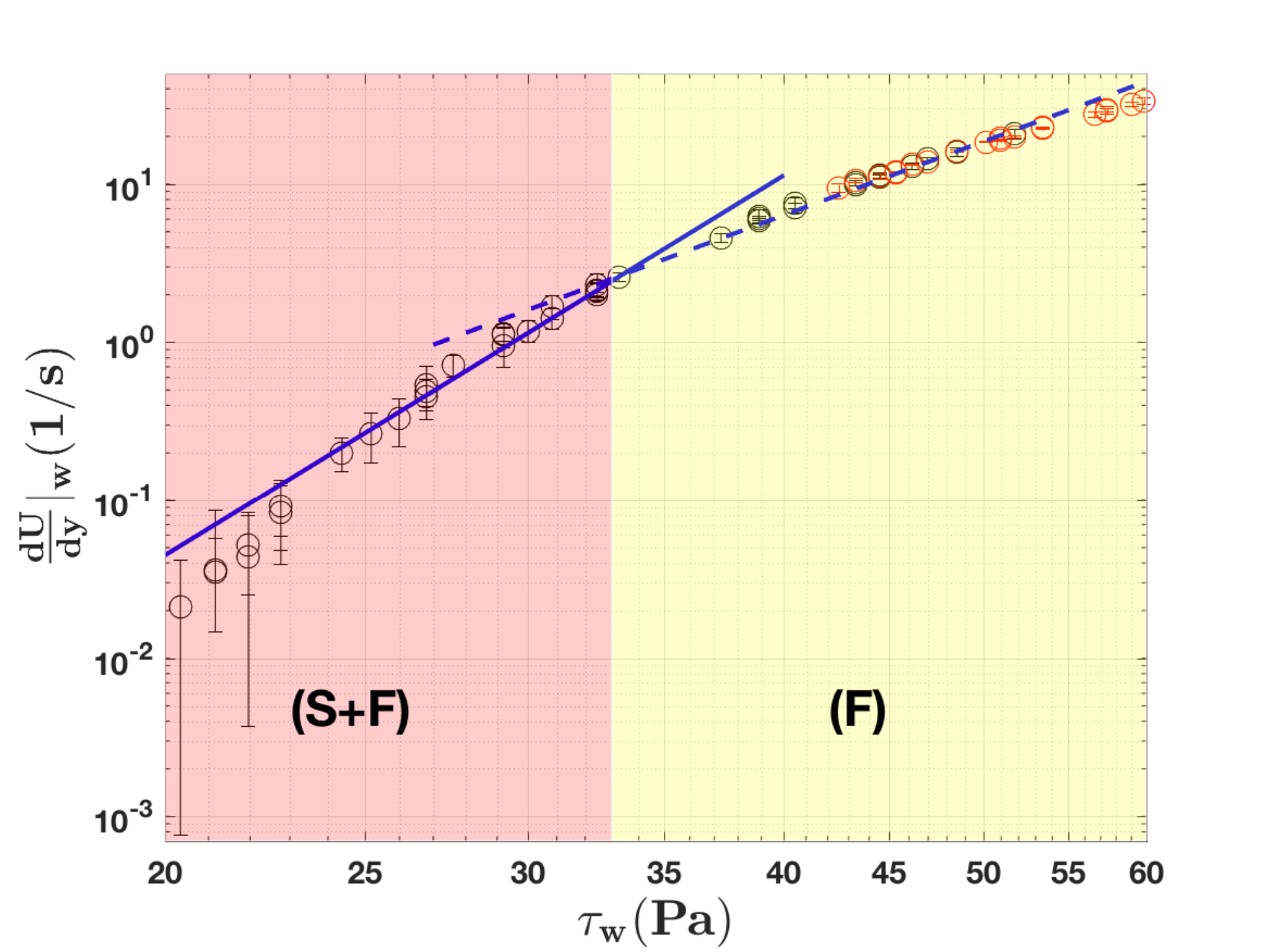}
   \label{slope_stress}
}
\caption{\subref{Up_Q} Dependence of the slip velocity $U_s$ on the volumetric flow rate $Q$. \subref{Us_Q} Dependence of the plug velocity $U_p$ on the volumetric flow rate $Q$. \subref{slope_stress} dependence of the wall velocity gradient $\frac{dU}{dy} \vert _w$ vs. wall shear stress $\tau_w$. Black/red circles correspond to the treated/untreated PMMA micro-channel.}
\label{F:UsUpdU}
\end{figure}

The comparison of the dependencies of the plug velocity $U_p$, slip velocity $U_s$ and the wall shear rates $\frac{dU}{dy} \vert _w$ on the driving flow rate $Q$ measured in chemically treated and untreated channels is shown in Fig. \ref{F:UsUpdU}. As already noted, for the case of the chemically treated channel the slip velocity is smaller than in the case of the untreated channel,  Fig. \ref{Us_Q}. Consequently, the plug velocity is systematically larger in the treated channel.

The dependence of the wall velocity gradients on the wall shear stress measured in the chemically treated channel is fully consistent with the measurements performed with the untreated channel in the third flow regime,  \ref{slope_stress}. This reinforces the statement that in the chemically treated channel a single flow regime is observed corresponding to the largest flow rates investigated with the untreated channel.

To gain further insights into the physical mechanism by which the chemical treatment affects the slip behavior we turn our attention to the direct visualisation of the flow in both untreated and chemically treated channels at the same flow rate $Q= 1 ~ \mu l/min$. Two distinct methods of flow visualisation have been employed. First, we have performed direct imaging of the streaks of the micro-PIV seeding particles by overexposing the images prior to acquisition. This kind of visualisation offers primary insights into the structure of the microscopic flow. Second, we display the corresponding microscopic flow fields. 

\begin{figure}
\centering
\includegraphics[height=6cm]{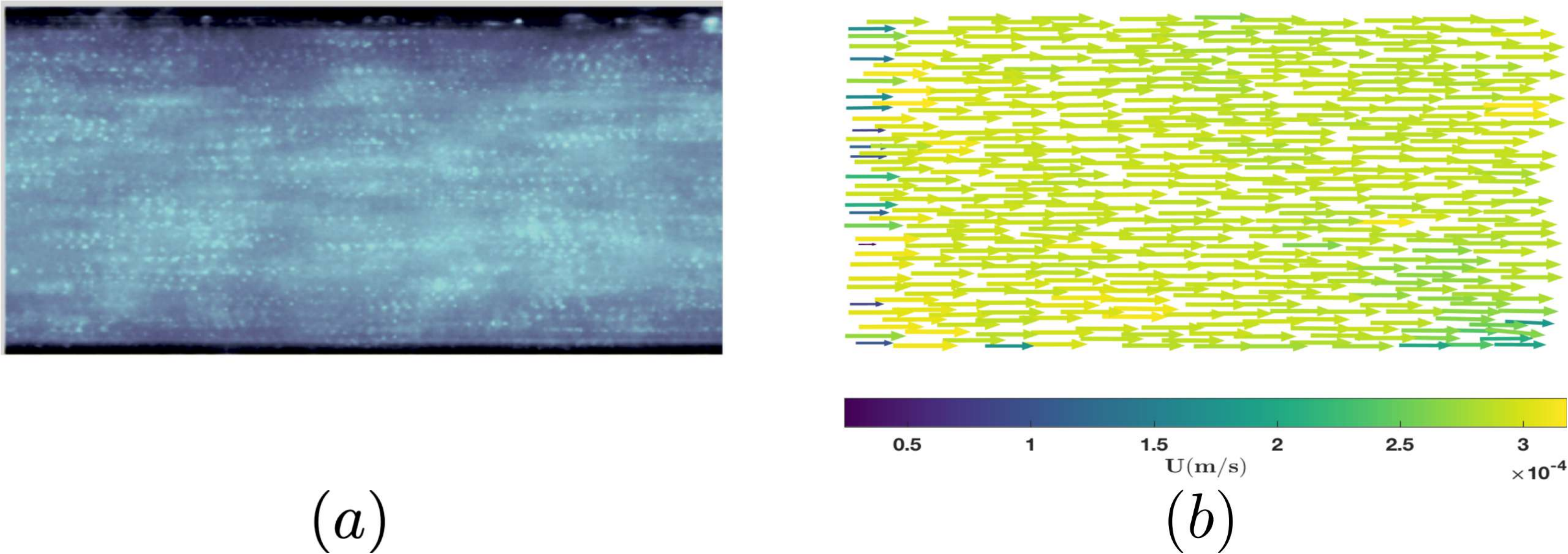}
\caption{(a) Streak imaging of the flow in the  untreated micro-channel. (b) Velocity field measured in the chemically untreated channel. The flow rate was $Q=1 ~\mu l/min$.
\label{F:Imaging_SLIP}}
\end{figure}

The flow in the untreated channel is described in Fig. \ref{F:Imaging_SLIP}. As one would expect for the case of a steady and laminar flow, the streak lines are rectilinear and parallel to the walls of the micro-channel, Fig. \ref{F:Imaging_SLIP}(a). Consequently, the velocity field exhibits no component in the direction transversal to the flow, Fig. \ref{F:Imaging_SLIP}(b). 

\begin{figure}
\centering
\includegraphics[height=6cm]{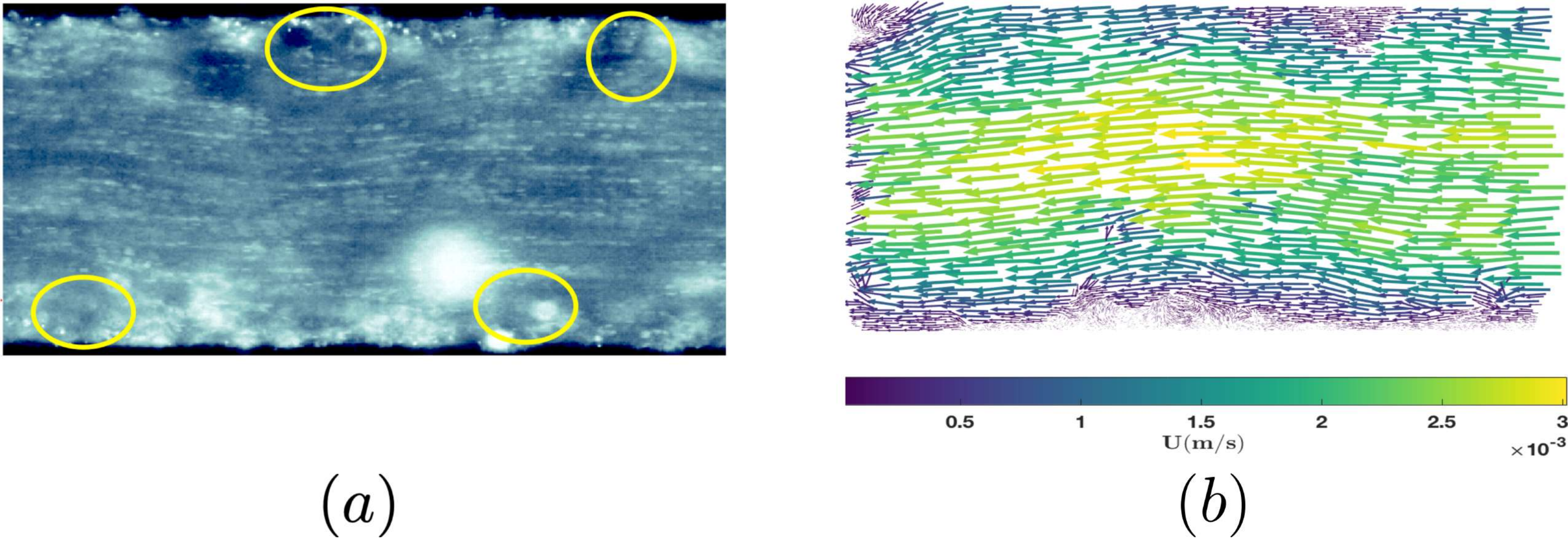}
\caption{(a) Streak imaging of the flow in the chemically treated micro-channel.  The yellow ellipses highlight blobs of the gel rigidly attached to the walls. (b) Velocity field measured in the chemically treated channel. The flow rate was $Q=1 \mu l/min$. \label{F:Imaging_NO_SLIP}}
\end{figure}

The flow in the chemically treated micro-channel is illustrated in Fig. \ref{F:Imaging_NO_SLIP}. The image of the streaks of the seeding particles reveal the presence of rigid blobs of Carbopol unevenly attached to the walls of the micro-channel (the regions highlighted by the yellow ellipses in Fig. \ref{F:Imaging_NO_SLIP}(a).) By monitoring the streak line behavior at various positions downstream we have observed no regularity of their distribution along the channel walls and thus we have concluded that their appearance is rather random. We attribute this apparent randomness to the polidispersity of the Carbopol microstructure which has been demonstrated experimentally, \cite{debruyn2,debruyn1}. The presence of these solid blobs attached to the micro-channel walls induces a secondary flow, Fig.  \ref{F:Imaging_NO_SLIP}(b). Whereas in a macroscopic geometry this effect might be of little or no importance, in a micro-channel with a size of the same order of magnitude as the average size of the gel particles this effect can no longer be ignored.

A second issue related to the overall effectiveness of this chemical treatment is its temporal stability. To address this issue we have performed macro-rheological tests similar to those presented in Fig. \ref{F:slip_effect} after keeping the chemically treated acrylic blocks in a beaker filled with deionized water for one week in conditions of gentle stirring with a magnetic stirring rod.  The result was that the slip behavior illustrated in Fig. \ref{F:slip_effect} was restored which indicated that the effectiveness of the treatment was basically lost - the squares in Fig. \ref{F:slip_effect}.

\section{Summary of conclusions, outlook}\label{sec:conclusion}
An experimental investigation of the wall slip phenomenon during a steady micro-flow of a Carbopol gel was carried out in this study. By means of Digital Particle Image Velocimetry, accurate measurements of the flow fields have been performed in a wide range of flow rates. Depending on the magnitude of the flow rate, three distinct flow regimes are observed. At low driving flow rates, a full plug flow regime is observed, Fig. \ref{F:Plug}.  The full plug flow regime may be understood in terms of a solid body of un-yielded gel sliding over a thin layer of depleted solvent expelled from the spongy Carbopol structural units in the vicinity of the solid walls.
As the flow rate is gradually increased, the Carbopol gel is partially yielded in the proximity of the channel walls but a central un-yielded plug may still be observed around the centre-line of the micro-channel, Fig. \ref{F:Transition}. As the flow rate is increased even further, the Carbopol gel is fully yielded across the entire width of the channel, Fig. \ref{F:fluid}. 

 As the flow rate is increased past the full plug regime, the plug length $W_p$ decreases monotonically from a value equal to half channel width $W/2$ to zero, Fig. \ref{F:Plug radius}.
The scaling behavior of the wall velocity gradients   $\frac{dU}{dy} \vert _w$ with the flow rate is illustrated in Fig. \ref{F:dU_vs_Q}. Within the partially yielded regime, $\frac{dU}{dy} \vert _w \propto Q^2$ whereas within the fully yielded regime $\frac{dU}{dy} \vert _w \propto Q^3$. 

Measurements of the velocity profiles allow one to obtain the wall velocity gradients and slip velocities by extrapolating the velocity profiles at the channel walls. We use the dependence between the stress and the rate of shear obtained via classical rotational rheometry and we relate the measured wall velocity gradients to the wall stresses in order to obtain experimentally the scaling laws for the wall slip phenomenon. 

As compared to a number of previous studies that inferred the wall velocity gradients and the wall shear stresses directly from the driving pressure drops and flow rates \cite{Geraud2013, RheoPIV,barentin}, this procedure has the important advantage of making absolutely no assumption on either the flow structure or the rheological behavior of the material.
 The scaling behavior of the wall velocity gradients with respect to the wall shear stresses is illustrated in Fig. \ref{F:dU_vs_shearstress}. Within the partially yielded regime, the wall velocity gradients scale as $\frac{dU}{dy} \vert _w \propto \tau_w^{7.99}$ whereas in the fully yielded regime as $\frac{dU}{dy} \vert _w \propto \tau_w^{4.8}$. 
 
 Two distinct scaling laws of the slip velocity $U_s$  with the wall shear stress $\tau_w$ are observed when the flow rates are increased past the full plug regime. Within the partially yielded flow regime, a power law scaling of the slip velocity with the wall shear stress  is observed, $U_s  \propto \tau_w^{3.55}$, the full line in Fig. \ref{F:Us_vs_shear}. Within the fully yielded flow regime, the scaling becomes linear, $U_s  \propto \tau_w$, the dashed line Fig. \ref{F:Us_vs_shear}. For both the partially yielded and fully yielded regimes, a power law scaling of the slip velocity with the wall velocity gradients $\frac{dU}{dy} \vert _w$ is observed, \ref{F:Us_vs_dU}.

Through the last part of the paper, we have investigated the effectiveness of a chemical treatment proposed by M\'{e}tivier and coworkers in ref. \cite{StickSlip} in removing the wall slip effects. Whereas the treatment is indeed able to remove the wall slip effects during macro-rheological tests, Fig. \ref{F:slip_effect}, its effectiveness at a microscopic scale is somewhat more limited. First, although the treatment is able to inhibit the wall slip effects (the slip velocities are smaller and the wall velocity gradients significantly larger), it does not completely remove them, Fig. \ref{F:Fluid_noslip} .
A second drawback of the chemical treatment relates to the emergence of secondary flow due to the presence of solid blobs attached to the walls of the micro-channel at positions randomly distributed downstream, Fig. \ref{F:Imaging_NO_SLIP}.  Third, the temporal stability of this treatment is limited: after a week, the effects of the treatment vanished and the macroscopic slip behavior was fully restored, the squares in Fig. \ref{F:slip_effect}.

The results presented in this study clearly call for both along several lines. 
On the experimental side, future studies that could capture the evolution of the micro-structure of the yield material with the flow would help clarifying the generally accepted (but, to our  best knowledge, not demonstrated experimentally) picture of a thin depleted layer of solvent in the vicinity of the solid surface. This is a rather difficult task as it requires the development of a chemical reaction able to graft a low molar mass fluorescent compound on the micro-structural units of the gel without altering its yield stress behavior. 

On a theoretical line, in order to understand the physical mechanisms  of  the wall slip, one needs to develop a theoretical physico-chemical framework able to confirm the formation of a depleted layer of the solvent near the wall which in turn would allow one to predict analytical laws for the scaling of the slip velocity with the wall shear stresses. Last, such physical picture of the wall slip needs to be later be incorporated in a novel constitutive law able to fully describe the flow in the presence of slip.


\section{Acknowledgements}
We acknowledge the Agence Nationale de la Recherche (ANR) for the financial support via project NaiMYS (ANR-16-CE06-0003). 

V.B. gratefully acknowledges a visiting fellowship from Polytech Nantes.

E.Y. and T.B. are deeply indebted to Prof. Georgios C. Georgiou for  valuable discussions on flows of yield stress fluids in the presence wall slip.

\bibliographystyle{aapmrev4-2}
\bibliography{Refs_GLOBAL.bib}

\end{document}